\documentclass[twocolumn,preprintnumbers,amsmath,amssymb,pra]{revtex4}

\usepackage{graphicx,amsmath}
\usepackage{dcolumn}
\usepackage{bm}
\usepackage{amssymb}
\usepackage{epstopdf}
\usepackage{color}
\graphicspath{{Figures/}}

\begin{document}

\title{Single-photon scattering on a qubit. Space-time structure of the scattered
field}

\begin{abstract}
We study the space-time structure of the scattered field induced
by the scattering of a narrow single-photon Gaussian pulse on a
qubit embedded in 1D open waveguide. For a weak excitation power
we obtain explicit analytical expressions for space and time
dependence of reflected and transmitted fields which are, in
general, are different from plain travelling waves. The scattered
field consists of two parts: a damping part which represent a
spontaneous decay of the excited qubit and a coherent, lossless
part. We show that for large distance $x$ from qubit and at times
$t$ long after the scattering event our theory provides the result
which is well known from the stationary photon transport. However,
the approach to the stationary limit is very slow. The scattered
field decreases as the inverse powers of $x$ and $t$ as both the
distance from the qubit and the time after the interaction
increase.
\end{abstract}

\pacs{84.40.Az,~ 84.40.Dc,~ 85.25.Hv,~ 42.50.Dv,~42.50.Pq}
 \keywords      {qubits, microwave circuits,
waveguide, transmission line, quantum measurements}

\date{\today}

\author{Ya. S. Greenberg}\email{yakovgreenberg@yahoo.com}
\affiliation{Novosibirsk State Technical University, Novosibirsk,
Russia}
\author{A. G. Moiseev} \affiliation{Novosibirsk State
Technical University, Novosibirsk, Russia}
\author{A. A. Shtygashev} \affiliation{Novosibirsk State
Technical University, Novosibirsk, Russia}


 \maketitle

\section{Introduction}

Manipulating the propagation of photons in a one-dimensional
waveguide coupled to an array of two-level atoms (qubits) may have
important applications in quantum devices and quantum information
technologies \cite{Rai2001, Roy2017, Gu2017}.

A single photon scattered by a single atom embedded in a 1D open
waveguide was first considered in \cite{Shen2005a, Shen2005b},
where the authors employed the real 1D space description of the
Dicke Hamiltonian and the Bethe-ansatz approach \cite{Rup1984} to
derive the stationary solution for the photon transport. It was
found that a photon with a frequency equal to that of the
two-level atom can be completely reflected due to quantum
interference. This property has been experimentally confirmed in
the scattering of a microwave photon by a superconducting qubit
\cite{Asta2010, Hoi2011, Hoi2013}.

Since then, theoretical calculations of the stationary photon
transport in a 1D open waveguide with the atoms placed inside have
been performed in a configuration space
\cite{Shen2009,Cheng2017,Fang2014,Zheng2013} or by alternative
methods such as those based on Lippmann-Schwinger scattering
theory \cite{Roy2011, Huang2013, Diaz2015}, the input-output
formalism \cite{Fan2010, Lal2013, Kii2019}, the non-Hermitian
Hamiltonian \cite{Green2015}, and the matrix methods
\cite{Green2021,Tsoi2008}.

Even though the stationary theory of the photon transport provides
a useful guide to what one would expect in real experiment, it
does not allow for a description of the dynamics of a qubit
excitation and the evolution of a single-photon pulse.

Within the framework of the stationary scattering theories there
are only incident and reflected plane waves in front of the qubit,
and the transmitted plane wave behind the qubit. The reflected and
transmitted amplitudes should be understood as the limits of
time-dependent description when both the time after the scattering
event and the distance from the qubit tend to infinity.  Within
this approach, all information about the temporal and spatial
evolution of the field scattered by the qubit is completely lost.
To obtain this information, it is necessary to consider a
time-dependent problem, when the incident wave is a wave packet
that depends on time and coordinates.

In practice, the qubits are excited by the photon pulses with
finite duration and finite bandwidth. Therefore, to study the real
time evolution of the photon transport and atomic excitation the
time-dependent dynamical theories were developed \cite{Chen2011,
Liao2015, Liao2016a, Liao2016b, Zhou2022}. In these works, the
dynamics of the amplitudes of the qubit, transmitted, and
reflected waves were considered with an incident single-photon
Gaussian packet being scattered by the qubit. The main attention
was paid to the reflected and transmitted spectra as the time
after the scattering event tends to infinity. In this case, the
field scattered by the qubit become plain waves and asymptotically
approaches the stationary results for the photon transport.

Even though the time-dependent theory allows, in principle, to
study the real-time evolution of the scattered field, the
systematic and exhaustive discussion of this issue  is absent
except for several numerical plots
\cite{Chen2011,Liao2015,Drob2000}. The investigation of the
electric field induced by the propagation of a single-photon wave
packet through a single atom embedded in a 1D waveguide has been
performed in \cite{Dom2002}. However, in this paper the frequency
dependence of transmitted and reflected fields has not been
studied. The main attention was paid to the on-resonance
dependance of transmittance and reflectance on the pulse width.

In the real case, the measurements are being performed shortly
after the qubit excitation. Under these conditions the reflected
and transmitted fields are not plain waves. Therefore, from the
point of device applications, it is very important to study a
real-time evolution and space structure of the scattered field.

In the present paper we consider the scattering of a narrow
single-photon Gaussian pulse by a two-level artificial atom
(qubit) embedded in an 1D open waveguide. We assume that the
bandwidth of the pulse is much smaller than that of any other
components of the system. It allows us to obtain the explicit
analytical expressions for the scattered waveguide fields. The
scattered fields consists of two parts: a damping part which
represents a spontaneous decay of the excited qubit and a
coherent, lossless part. We show that for large distance $x$ from
qubit and at times $t$ long after the scattering event our theory
for the reflected and transmitted amplitudes provides the result
which is well known from the stationary scattering theories.
However, in general, the structure of the scattered field is
different from the stationary limit.

The paper is organized as follows. In Sec. II we introduce the
basic parameters describing the transmitted and reflected
amplitudes and their asymptotic properties. A general description
of our model is given in Sec. III. The interaction between the
qubit and electromagnetic field is described by Jaynes-Cummings
Hamiltonian. The trial wave-function is taken within a
single-excitation subspace. From time-dependent Schrodinger
equation we obtain the single-photon amplitudes for forward and
backward waves. The main result of the paper is given in Sec. IV.
Here we construct the photon wavepacket for forward and backward
propagating fields. We obtain the explicit expressions for the
functions $F_{T(R)} (\omega_S ,x,t)$ which are given in equations
(\ref{I3}), (\ref{I4}). We show that as both a distance from the
qubit and the time after scattering tend to infinity, the
stationary results (\ref{I1}), (\ref{I2}) are recovered. The
details of the calculations are given in Appendices A and B. The
influence of the probing power, decoherence rate, and the non
radiative losses on the transmitted and reflected fields are
explained in Appendix C.

\section{Formulation of the problem}

\begin{figure}
  \includegraphics[width=8 cm]{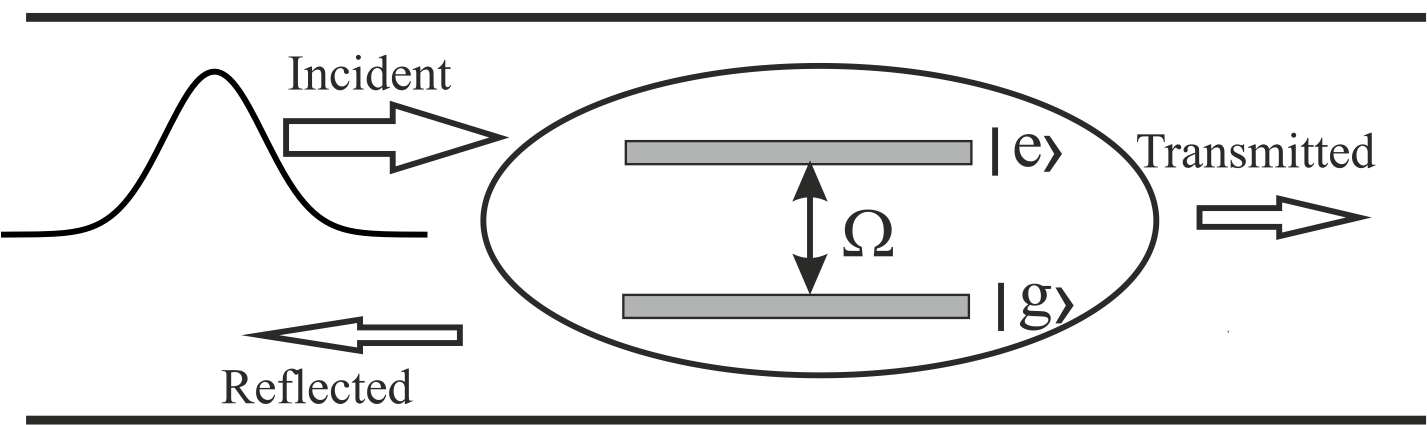}\\
  \caption{Schematic representation of a single-photon Gaussian pulse
  interacting with a two-level atom with energy levels $|g\rangle$ and $|e\rangle$,
respectively. $\Omega$ is the separation between the energy
levels. Long horizontal lines denote the waveguide
geometry.}\label{Fig0}
\end{figure}

We consider the interaction of a single-photon Gaussian pulse with
a two-level atom which is coupled to the waveguide modes with a
strength $g$. The excitation frequency of a qubit is $\Omega$ (see
Fig. \ref{Fig0}). A qubit is considered as the point-like emitter
which is placed in the point $x=0$ of the $x$ axis. We assume the
interaction of incident pulse with the qubit starts at $t=0$. It
results in the reflected and transmitted fields whose space and
time structure is the main subject of the paper.

Even though our treatment can be applied to a real two-level atom,
we consider here an artificial two-level atom, a superconducting
qubit operating at microwave frequencies. For subsequent
calculations we take typical qubit's parameters: the excitation
frequency, $\Omega/2\pi=5$ GHz which corresponds to the wavelength
$\lambda=6$ cm, the rate of spontaneous emission into waveguide
modes, $\Gamma/2\pi=10$ MHz. We asssume the group velocity of
electromagnetic waves is equal to that of a free space,
$v_g=3\times 10^8$ m/s.

The analytical expressions for the transmission $T$ and reflection
$R$ amplitudes found in a framework of stationary scattering
approach for a monochromatic signal scattered by a two-level atom
in an 1D open waveguide \cite{Shen2005a,Shen2005b} are as follows:

\begin{equation}\label{I1}
T(\omega_S) = \frac{{\omega _s  - \Omega }} {{\omega _s  - \Omega
+ i\frac{\Gamma}{2} }}
\end{equation}
\begin{equation}\label{I2}
R(\omega_S) = \frac{{ - i\frac{\Gamma}{2} }} {{\left( {\omega _s -
\Omega + i\frac{\Gamma}{2} } \right)}}
\end{equation}

where $\omega_s$ is the photon frequency.

It follows from expression (\ref{I1}) that when the photon
frequency coincides with the qubit frequency, then the value of T
vanishes. In this case, the incident photon is completely
reflected from the qubit.  The reason for this perfect reflection
is a coherent interference between the incident wave and the wave
scattered  by the qubit. It can be said that in this case, the
qubit plays the role of an ideal mirror. This behavior was first
observed experimentally in the scattering of a microwave photon by
a superconducting qubit \cite{Asta2010}.

Strictly speaking the  equations (\ref{I1}), (\ref{I2}) are valid
if we assume a weak probing signal and neglect qubit's pure
dephasing, $\Gamma_\varphi$ and non-radiative intrinsic losses,
$\Gamma_l$ \cite{Hoi2013}. In our treatment below we assume the
probe is weak. Under this assumption the pure dephasing and
non-radiative losses can simply be incorporated in our treatment
by adding the imaginary part to the qubit's frequency $\Omega$,
$\Omega\rightarrow\Omega-i(\Gamma_\varphi+\Gamma_l/2)$. We address
this issue in a more detail in the Appendix C.

From general considerations, it is obvious that the plane wave
solutions (\ref{I1}), (\ref{I2}) should be a limiting case of a
time-dependent picture when both the distance from a qubit and the
time after the scattering tend to infinity.  Near the qubit, the
scattered field is more complicated, the amplitude of which
depends on the space-time coordinates x, t of the scattered field.
In the general case, the transmission and reflection fields should
have the following form:
\begin{equation}\label{I3}
T(\omega_S,x,t) = T(\omega_S)e^{i\frac{\omega_S}{vg}(x-v_gt)} +
F_T (\omega_S ,x,t)
\end{equation}

\begin{equation}\label{I4}
R(\omega_S,x,t) = R(\omega_S)e^{-i\frac{\omega_S}{vg}(x+v_gt)} +
F_R (\omega_S ,x,t)
\end{equation}

The quantities $F_{T(R)} (\omega_S ,x,t)$ that characterize the
space-time structure of the scattered field must satisfy the
following property: $F_{T(R)} (\omega_S ,x,t) \to 0 $ at $\left| x
\right| \to \infty ,\;t \to \infty$. Their structure depends, of
course, on the shape of the initial pulse. For a Gaussian wave
packet, the structure of  $F_{T(R)} (\omega_S ,x,t)$  can be
studied only by numerical methods \cite{Chen2011, Liao2015}.

In the present paper, we take the excitation pulse in the form of
a Gaussian wave packet which is given by \cite{Liao2015}:

\begin{equation}\label{G1}
\gamma _k (0) = \left( {\frac{{8\pi }} {{L^2 \Delta _k^2 }}}
\right)^{1/4} e^{ - \frac{{(k - k_S )^2 }} {{\Delta _k^2 }}}
\end{equation}

where $L$ is the length of a waveguide, $k_S$ is the wave vector
corresponding to the center frequency of the pulse, and $\Delta_k$
is the width in the k space. We note that the pulse (\ref{G1})
 ensures that there is only a single photon in the wavepacket:
 $\frac{L}
{{2\pi }}\int\limits_{ - \infty }^\infty  {\left| {\gamma _k (0)}
\right|^2 dk}  = 1$.

For narrow pulse we obtain the explicit analytical expressions for
the functions $ F_{T(R)} (\omega_S ,x,t)$. These functions
decrease relatively slow (as the inverse powers of x and t) for
large both $x$ and $t$, however, for $\;t \to \infty$ and fixed
$x$, these functions do not tend to zero. It means that at
relatively small distances from the qubit, the field is not
uniform and the dependence of the transmission and reflection
amplitudes on the frequency is more complicated than it follows
from expressions (\ref{I1}), (\ref{I2}).

\section{The model}
We consider a single qubit which is located at the point $x=0$ in
an open linear waveguide.  The Hilbert space of the qubit consists
of the excited state $|e\rangle$, and the ground state
$|g\rangle$. The Hamiltonian which accounts for the interaction
between the qubit and electromagnetic field is as follows (we use
units where $\hbar=1$ throughout the paper):

\begin{equation}\label{1}
\begin{gathered}
H = H_0  + \sum\limits_{k>0} {\omega _k a_k^ +  a_k } + \sum\limits_{k<0}{\omega _k b_k^ +  b_k }  \hfill \\
+ \sum\limits_{k>0} {\left( {g_k^{} \sigma _ - ^{} a_k^ +   +
g_k^{} \sigma _ + ^{} a_k^{} } \right)}
+ \sum\limits_{k<0} {\left( {g_k^{} \sigma _ - ^{} b_k^ +   + g_k^{} \sigma _ + ^{} b_k^{} } \right)}  \hfill \\
\end{gathered}
\end{equation}

where $H_0$ is Hamiltonian of  bare qubit,

\begin{equation}\label{2}
H_0  = \frac{1} {2}\left( {1 + \sigma _z^{} } \right)\Omega
\end{equation}

The photon operators $a^+_k$, $a_k$ ($k>0$), and $b^+_k$, $b_k$
($k<0$) describe forward and backward scattering waves,
respectively; $\sigma_+$, $\sigma_-$ are the rising and lowering
spin operators, respectively: $\sigma_+=|e\rangle\langle g|$,
$\sigma_-=|g\rangle\langle e|$. A spin operator
$\sigma_z=|e\rangle\langle e|-|g\rangle\langle g|$. The quantity
$g_k$ in (\ref{1}) is the coupling between qubit and the photon
field in a waveguide. We assume that the coupling is the same for
forward and backward waves, $g_k=g_{-k}$.

Below we consider a single-excitation subspace with either a
single photon being in a waveguide and the qubit being in the
ground state, or there are no photons in a waveguide with the
qubit being excited. Therefore, we limit Hilbert space to the
following states:
\begin{equation}\label{3}
\begin{gathered}
  \left| {g,0} \right\rangle  = \left| g \right\rangle  \otimes \left| 0 \right\rangle  \hfill \\
  a_k^ +  \left| {g,0} \right\rangle  = \left| g \right\rangle  \otimes a_k^ +  \left| 0 \right\rangle  \hfill \\
  b_k^ +  \left| {g,0} \right\rangle  = \left| g \right\rangle  \otimes b_k^ +  \left| 0 \right\rangle  \hfill \\
\end{gathered}
\end{equation}
A trial wave function in single-excitation subspace reads:

\begin{equation}\label{4}
\begin{gathered}
  \left| \Psi  \right\rangle  = \beta (t)e^{ - i\Omega t} \left| {e,0} \right\rangle  + \sum\limits_{k>0} {\gamma _k (t)e^{ - i\omega _k t} } a_k^ +  \left| {g,0} \right\rangle  \hfill \\
   + \sum\limits_{k<0} {\delta _k (t)e^{ - i\omega _k t} b_k^ +  } \left| {g,0} \right\rangle  \hfill \\
\end{gathered}
\end{equation}

where $\beta(t)$  is the amplitude of the qubit, $\gamma_k(t)$,
and $\delta_k(t)$ are the single-photon amplitudes for forward and
backward waves, respectively.

The equations for the quantities  $\beta(t)$, $\gamma_k(t)$, and
$\delta_k(t)$  can be found from time-dependent Schrodinger
equation $id|\Psi\rangle/dt=H|\Psi\rangle$.

\begin{equation}\label{5}
\frac{{d\beta }} {{dt}} =  - i\sum\limits_{k>0} {} g_k^{} \gamma
_k(t) e^{ - i(\omega _k  - \Omega )t}  - i\sum\limits_{k<0} {}
g_k^{} \delta _k(t) e^{ - i(\omega _k  - \Omega )t}
\end{equation}

\begin{equation}\label{6}
\frac{{d\gamma _k }} {{dt}} =  - i\beta(t) g_k e^{i(\omega _k  -
\Omega )t}
\end{equation}

\begin{equation}\label{7}
\frac{{d\delta _k }} {{dt}} =  - i\beta(t) g_k e^{i(\omega _k  -
\Omega )t}
\end{equation}
From equations (\ref{6}) and (\ref{7}) we obtain:

\begin{equation}\label{8}
\gamma _k (t) = \gamma _k (0 ) - ig_k \int\limits_0^t {} \beta
(t')e^{i(\omega _k  - \Omega )t'} dt'
\end{equation}

\begin{equation}\label{9}
\delta _k (t) =  - ig_k \int\limits_0^t {} \beta (t')e^{i(\omega
_k  - \Omega )t'} dt'
\end{equation}

Substitution of (\ref{9}) and (\ref{8}) into equation (\ref{5})
and application of Wigner-Weisskopf approximation provide the
following expression for the qubit amplitude $\beta(t)$ (the
details of the derivation are given in Appendix A):

\begin{equation}\label{10}
\frac{{d\beta }} {{dt}} =  - i\frac{\sqrt{L\Gamma}} {\sqrt{2\pi^2}
} \int\limits_0^{  \infty } {\gamma _0 (\omega )e^{ - i(\omega -
\Omega )t} d\omega }  - \frac{\Gamma}{2} \beta
\end{equation}

where
\begin{equation}\label{G2}
\gamma _0 (\omega ) = \left( {\frac{{8\pi }} {{L^2 \Delta ^2 }}}
\right)^{1/4} e^{ - \frac{{(\omega  - \omega _S )^2 }} {{\Delta ^2
}}},
\end{equation}

\begin{equation}\label{12}
\Gamma  = 2\pi \sum\limits_k {g_k^2 \delta (\omega _k  - \Omega )}
\end{equation}

The form of $\gamma_0(\omega)$ insures the validity of a
single-photon approximation: $\frac{L} {{2\pi }}\int\limits_{ -
\infty }^\infty  {\left| {\gamma _0 (\omega )} \right|^2 d\omega }
= 1$.

The integral in the righthand side of Eq.\ref{10} can be expressed
in terms of the error function $\texttt{erf}(x)$ \cite{Grad2007}:
\begin{equation}\label{G3}
\begin{gathered}
  \int\limits_0^\infty  {} \gamma _0 (\omega )e^{ - i(\omega  - \Omega )t} d\omega  \hfill \\
   = 2^{-1/4}\pi^{3/4} \sqrt{\frac{\Delta }
{L}}e^{ - \frac{{\Delta ^2 t^2 }} {4}} \left[ {1 -
\texttt{erf}\left( {it\frac{\Delta } {2} - \frac{{\omega _S }}
{\Delta }} \right)} \right]e^{ - i(\omega _S  - \Omega )t}  \hfill \\
\end{gathered}
\end{equation}
From now on we consider a narrow pulse where $\Delta$ is a small
quantity. In the leading order in $\Delta$ we obtain from
(\ref{G3}):
\begin{equation}\label{G4}
\int\limits_0^\infty  {} \gamma _0 (\omega )e^{ - i(\omega  -
\Omega )t} d\omega  = (2\pi) ^{3/4} \sqrt {\frac{\Delta } {L}} e^{
- i(\omega _S  - \Omega )t}
\end{equation}
This approximation is valid for $\Delta \ll \omega_S$, $\Delta
t\ll 1$.

Regarding the Eq. \ref{G4}, it is worth noting that in our case a
narrow Gaussian pulse can be approximated by a delta pulse with
the amplitude $(2\pi) ^{3/4} \sqrt {\frac{\Delta }{L}}$ :

 \begin{equation}\label{G4a}
    \gamma_0(\omega)=(2\pi) ^{3/4} \sqrt {\frac{\Delta }
    {L}}\delta(\omega-\omega_S)
\end{equation}

Finally, the Eq.\ref{10} takes the form:
\begin{equation}\label{G5}
\frac{{d\beta }} {{dt}} =  - i\left(\frac{2}{\pi}\right)^{  1/4}
\sqrt {\Gamma \Delta } \;e^{-i(\omega _S  - \Omega )t}
-\frac{\Gamma}{2} \beta
\end{equation}

For initially unexcited qubit, $\beta(0)=0$, we obtain from
(\ref{10}) the following result for the qubit amplitude:

\begin{equation}\label{13}
\beta \left( t \right) = C_0 \left( {e^{ - \frac{\Gamma}{2} t}  -
e^{ - i(\omega _s  - \Omega )t} } \right)
\end{equation}
where
\begin{equation}\label{14}
C_0  =  -
\left(\frac{2}{\pi}\right)^{1/4}\frac{\sqrt{\Gamma\Delta}}{{\left(
{\omega _s - \Omega + i\frac{\Gamma}{2} } \right)}}
\end{equation}

Finally, for the forward- propagating wave, $\gamma_k(t)$,  with
$\beta(t)$ from (\ref{13}) we obtain:
\begin{equation}\label{15}
\gamma_k (t) = \gamma_k(0)+ \gamma _1 (\omega_k ,t)
\end{equation}
where
\begin{equation}\label{16}
\gamma _1 (\omega_k ,t) =  - g_k C_0 \left[ {I_1 (\omega_k ,t) -
iI_2 (\omega_k ,t)} \right],
\end{equation}

\begin{equation}\label{17}
I_1 (\omega_k ,t) = \frac{{\left( {e^{i\left( {\omega_k  - \Omega
+ i\frac{\Gamma}{2} } \right)t}  - 1} \right)}} {{\left( {\omega_k
- \Omega + i\frac{\Gamma}{2} } \right)}},
\end{equation}

\begin{equation}\label{18}
I_2 (\omega_k ,t) = \int\limits_0^t {dt'} e^{i\left( {\omega_k  -
\omega _s } \right)t'}  = \frac{{e^{i\left( {\omega_k  - \omega _s
} \right)t}  - 1}} {{^{i\left( {\omega_k  - \omega _s } \right)}
}}
\end{equation}

From (\ref{8}) and (\ref{9}) we may conclude that the amplitude of
the forward propagating (transmitted) wave is equal to the
amplitude of the backward propagating (reflected) wave,
$\delta_k(t)=\gamma_1(\omega_k,t)$.

As is pointed in Sect.II and is proved in the Appendix C our
treatment is valid if we consider a single-photon Gaussian pulse
as a weak excitation probe.  For single-photon transport the weak
excitation means that the pulse duration is much longer than the
spontaneous lifetime of the qubit, $\Delta\ll\Gamma$
\cite{Fan2010}. In this case, the qubit is mostly in the ground
state. Therefore, we can define the average number of probe
photons per interaction time $2\pi/\Delta$ as $N=2\pi
P/(\hbar\Omega\Delta)$, where $P$ is the power of the incident
pulse \cite{Hoi2011, Hoi2013}. Taking $\Delta/2\pi=1$ MHz,
$\Omega/2\pi=5$ GHz we can estimate the power of the incident
single-photon Gaussian probe in a weak excitation limit, $P\approx
\hbar\Omega\Delta/2\pi=4\times 10^{-17}$ W. This value is within a
reach of experimental technique \cite{Hoi2011, Hoi2013}.

\section{Space-time structure of the scattered field}
\subsection{Forward scattering field}
The photon wave packet for forward propagating field behind the
qubit is given by

\begin{equation}\label{Fw}
\begin{gathered}
  u(x,t) = \sum\limits_{k>0} {} \gamma _k (t)e^{i\frac{{\omega _k }}
{{v_g }}(x - v_g t)}  \hfill \\
   = \sum\limits_k {} \gamma _k (0)e^{i\frac{{\omega _k }}
{{v_g }}(x - v_g t)}  + \sum\limits_{k>0} {} \gamma _1 (\omega _k
,t)e^{i\frac{{\omega _k }}
{{v_g }}(x - v_g t)}  \hfill \\
\end{gathered}
\end{equation}
where $\gamma_k(0)$ is given in (\ref{G1}).

In equations (\ref{Fw}) $x>0$ and $x-v_gt<0$. The second condition
insures the causality of propagating field which appears at the
point $x$ behind the qubit not until the signal travels the
distance $x$ after the scattering.

For forward scattering the summation over $k$ is replaced by the
integration:

\begin{equation}\label{sum}
\sum\limits_{k>0} {}  \Rightarrow \frac{L}{{2\pi }}\int\limits_{ 0
}^\infty  {dk}  = \frac{{L}}{{2\pi \upsilon _g
}}\int\limits_0^\infty {d\omega }
\end{equation}
where we take a linear frequency dispersion $\omega=v_g|k|$ well
above the cutoff frequency of a waveguide.

The first sum in righthand side of (\ref{Fw}) reads:
\begin{equation}\label{sum1}
\begin{gathered}
  \sum\limits_k {} \gamma _k (0)e^{i\frac{{\omega _k }}
{{v_g }}(x - v_g t)}  = \frac{L} {{\pi \sqrt {v_g }
}}\int\limits_0^\infty  {} \gamma _0 (\omega )e^{i\frac{\omega }
{{v_g }}(x - v_g t)} d\omega  \hfill \\
   = \left( {\frac{8}
{\pi }} \right)^{1/4} \sqrt {\frac{{\Delta L}} {{v_g }}}
e^{i\frac{{\omega _S }}
{{v_g }}(x - v_g t)}  \hfill \\
\end{gathered}
\end{equation}
where we use a small $\Delta$ approximation (\ref{G4}). Therefore,
we may consider pre factor in (\ref{sum1}) as the amplitude of
incoming wave, $A=(8/\pi)^{1/4}\sqrt{\Delta L/v_g}$.

For the second sum in righthand side of (\ref{Fw}) we obtain:

\begin{equation}\label{sum2}
\begin{gathered}
  \sum\limits_{k>0} {} \gamma _1 (\omega _k ,t)e^{i\frac{{\omega _k }}
{{v_g }}(x - v_g t)}  =  - g_0 C_0 \frac{L}
{{2\pi v_g }}\left( {I_1 (x,t) - iI_2 (x,t)} \right) \hfill \\
   =A\frac{\Gamma }
{{4\pi }}\frac{1}{\omega_S-\Omega+i\frac{\Gamma}{2}}\left( {I_1 (x,t) - iI_2 (x,t)} \right) \hfill \\
\end{gathered}
\end{equation}

where

\begin{equation}\label{20a}
I_1 (x,t) = \int\limits_0^\infty  {} I_1 \left( {\omega ,t}
\right)e^{i\frac{\omega } {{v_g }}\left( {x - v_g t} \right)}
d\omega
\end{equation}

\begin{equation}\label{20b}
I_2 (x,t) = \int\limits_0^\infty  {} I_2 \left( {\omega ,t}
\right)e^{i\frac{\omega }
{{v_g }}\left( {x - v_g t} \right)} d\omega  \hfill \\
\end{equation}

In Eq. \ref{sum2} we use the on resonance value of the photon-
qubit coupling $g_0$, $g_0=\sqrt{\Gamma v_g/2L}$ (see (\ref{A8})
in Appendix A).

The calculations of the quantities $I_1(x,t)$, $I_2(x,t)$ are
performed in the Appendix B.  They are given by

\begin{equation}\label{21}
\begin{gathered}
  I_1 (x,t) =   e^{ - i\tilde \Omega \;t} e^{i\frac{x}
{{v_g }}\tilde \Omega } E_1\left({ - i\frac{x}
{{v_g }}\tilde \Omega } \right) \hfill \\
   + 2\pi ie^{i\frac{{\tilde \Omega }}
{{v_g }}(x - v_g t)}  - e^{ - i\frac{{\left| {x - v_g t} \right|}}
{{v_g }}\tilde \Omega } E_1\left(- {i\frac{{\left| {x - v_g t}
\right|}}
{{v_g }}\tilde \Omega } \right)\; \hfill \\
\end{gathered}
\end{equation}

\begin{equation}\label{22}
\begin{gathered}
  I_2 (x,t) = e^{i\frac{{\omega _s }}
{{v_g }}\left( {x - v_g t} \right)} \left( {2\pi  + i\,ci(\omega
_s \frac{x} {{v_g }}) + si(\omega _s \frac{x}
{{v_g }})} \right. \hfill \\
  \left. { - i\,ci(\omega _s \frac{{\left| {x - v_g t} \right|}}
{{v_g }}) + si(\omega _s \frac{{\left| {x - v_g t} \right|}}
{{v_g }})} \right) \hfill \\
\end{gathered}
\end{equation}

where $\widetilde{\Omega}=\Omega-i\frac{\Gamma}{2}$, $E_1(z)$ is
the exponential integral \cite{Abram1964}; $si(xy)$ and $ci(xy)$
are sine integral and cosine integral, respectively
\cite{Grad2007}.

\begin{equation}\label{23}
\begin{gathered}
  ci(xy) =  - \int\limits_x^\infty  {} \frac{{\cos zy}}
{{^z }}dz \hfill \\
  si(xy) =  - \int\limits_x^\infty  {} \frac{{\sin zy}}
{{^z }}dz \hfill \\
\end{gathered}
\end{equation}
where $y$ is $x/v_g$ or $|x-v_gt|/v_g$. In equations (\ref{21}),
(\ref{22}) $x>0,\quad x-v_gt<0$.

Combining (\ref{sum1}) and (\ref{sum2}) we obtain the forward
propagating field behind the qubit in the following form:

\begin{widetext}
\begin{equation}\label{23a}
\begin{gathered}
  u\left( {x,t} \right)/A = T(\omega_S)e^{i\frac{{\omega _s }}
{{v_g }}\left( {x - v_g t} \right)}  \hfill \\
   + \frac{iR(\omega_S)}{{2\pi }}e^{i\frac{{ \Omega }} {{v_g }}(x - v_g t)}e^{\frac{{\Gamma/2 }} {{v_g }}(x - v_g t)}
   \left(  E_1\left( {  i\frac{x} {{v_g }}\tilde \Omega }
\right) + 2\pi i - E_1\left( -{i\frac{{\left| {x - v_g t}
\right|}}
{{v_g }}\tilde \Omega } \right) \right) \hfill \\
   + \frac{R(\omega_S)}
{{2\pi }}e^{i\frac{{\omega _s }} {{v_g }}\left( {x - v_g t}
\right)} \left( {ici(\omega _s \frac{x} {{v_g }}) + \,si(\omega _s
\frac{x} {{v_g }}) - ici(\omega _s \frac{{\left| {x - v_g t}
\right|}} {{v_g }}) + si(\omega _s \frac{{\left| {x - v_g t}
\right|}}
{{v_g }})} \right) \hfill \\
\end{gathered}
\end{equation}

where $x>0$, $x-v_gt<0$.
\end{widetext}

It should be noted that the amplitude of transmitted field,
$T(\omega_S)$ in the first term in (\ref{23a}) is the result of a
summation of the incident wave  (\ref{sum1}) with a part of the
scattered field (2$\pi$ in (\ref{22})). The second term in
(\ref{23a}) is a damping part of the scattered field which
represents a spontaneous decay of the excited qubit. The third
term in (\ref{23a}) is a coherent, lossless part of the scattered
field. As both a distance from the qubit and the time after the
scattering tend to infinity these scattering fields die out
leaving only the plane wave stationary solution.

\begin{figure}
  \includegraphics[width=8 cm]{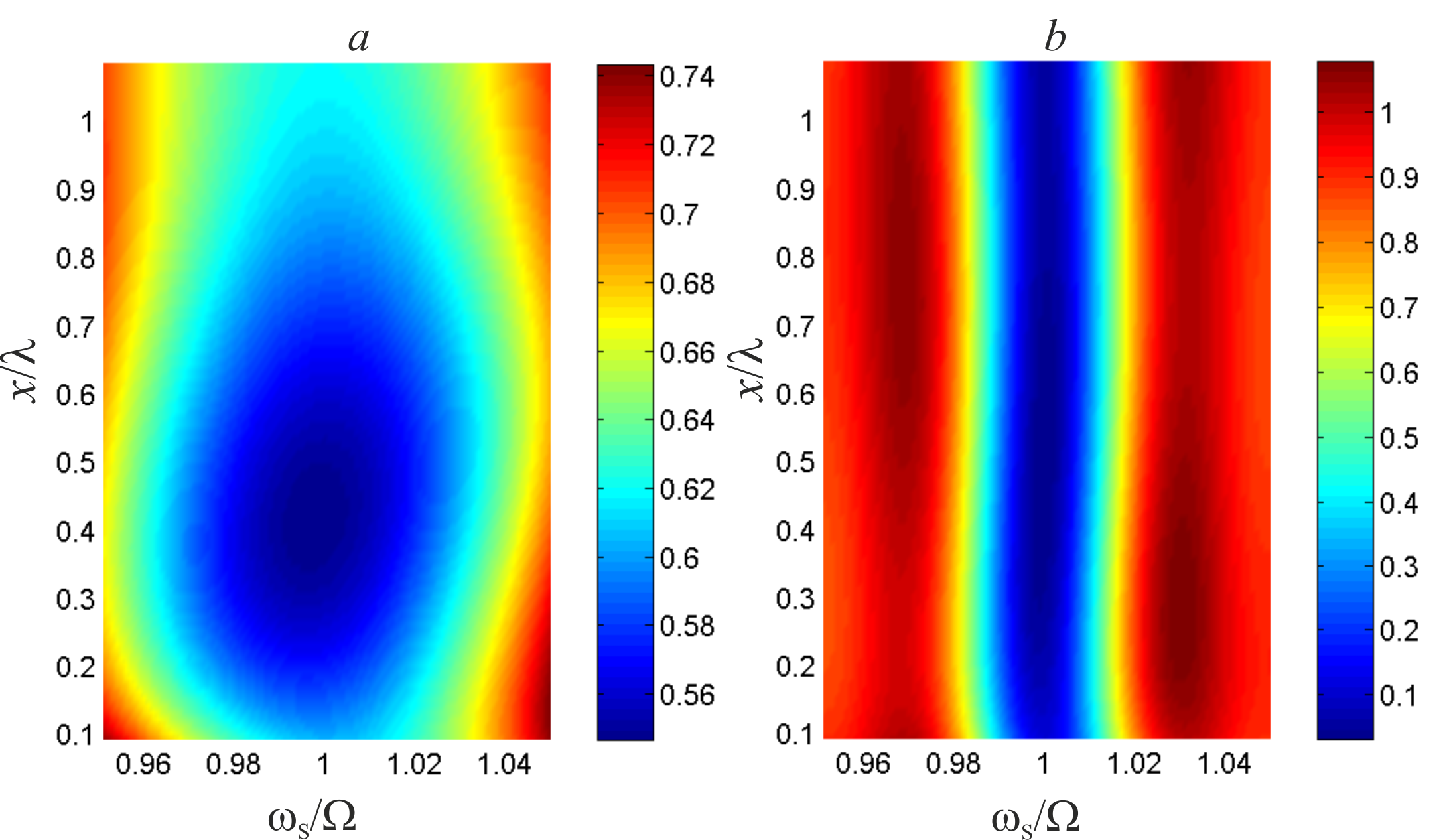}\\
  \caption{2D map of the transmittance calculated from (\ref{23a})
  for (a) $t=1 $ ns and (b) $t=5 $ ns. The color bar shows the value
  $|u(\omega_S,x,t)/A|^2$. $\Gamma/2\pi=0.01$ GHz, $\Omega/2\pi=5$ GHz,
   $\lambda=6$ cm.}\label{Fig2A}
\end{figure}

Two dimensional maps of the transmittance $|u(\omega_S,x,t)/A|^2$
calculated from (\ref{23a}) for $t=1 $ ns, $t=5 $ ns are shown in
Fig. \ref{Fig2A}.

In Fig.\ref{Fig2A}b we observe the off-resonance regions
($\omega_S/\Omega\approx 0,97, 1.03$) where the transmittance is
about $10\%$  larger than 1. This effect persists over wavelength
scale. It can be attributed to the interference between incident
wave and the field generating by the qubit itself. It does not
contradict with the conception of the probability. In relation to
our study the probability is inferred from the conservation of the
energy flux: at any instant of time the input energy flux is the
sum of the transmitted and reflected energy fluxes integrated over
all space and over all frequencies. In our paper we calculate not
the energy flux, but the electric field $u(x,t)$. Therefore, in this
case the conception of the probability is not applicable. As aside
comment it is worth noting that a similar amplification of the
field exists in Fabry-Perot interferometer with semi-transparent
mirrors \cite{Ley1987}.

\subsection{Backward scattering field}
The photon wave packet for backward propagating field before the
qubit is as follows:
\begin{equation}\label{24}
\begin{gathered}
 u(x,t)= \sum\limits_{k<0} {} \delta_k (t)e^{-i\frac{{\omega _k }}
{{v_g }}(x + v_g t)} \hfill \\ =  - g_0 C_0 \frac{L} {{2\pi v_g
}}\left( {J_1 (x,t) - iJ_2 (x,t)} \right)
\end{gathered}
\end{equation}

where
\begin{equation}\label{25a}
J_1 (x,t) = \int\limits_0^\infty  {} I_1 \left( {\omega ,t}
\right)e^{ - i\frac{\omega } {{v_g }}\left( {x + v_g t} \right)}
d\omega
\end{equation}

\begin{equation}\label{25b}
J_2 (x,t) = \int\limits_0^\infty  {} I_2 \left( {\omega ,t}
\right)e^{ - i\frac{\omega } {{v_g }}\left( {x + v_g t} \right)}
d\omega
\end{equation}

In equations (\ref{24}), (\ref{25a}), and (\ref{25b}) $x<0$ and
$x+v_gt>0$. The second condition insures the causality of the
backscattering field which appears at the point $x$ in front of
the qubit not until the signal travels the distance $|x|$ after
the scattering.

The quantities $J_1(x,t)$, $J_2(x,t)$ can be calculated similar to
the quantities $I_1(x,t)$, $I_2(x,t)$. The result is as follows:

\begin{equation}\label{26a}
\begin{gathered}
  J_1 (x,t) =  e^{ - i\tilde \Omega \;t} e^{i\frac{{\left| x \right|}}
{{v_g }}\tilde \Omega } E_1\left( {  i\frac{{\left| x \right|}}
{{v_g }}\tilde \Omega } \right) + 2\pi ie^{ - i\frac{{\tilde
\Omega }}
{{v_g }}(x + v_g t)}  \hfill \\
   - e^{ - i\frac{{x + v_g t}}
{{v_g }}\tilde \Omega } E_1\left( -{i\frac{{x + v_g t}}
{{v_g }}\tilde \Omega } \right)\; \hfill \\
\end{gathered}
\end{equation}

\begin{equation}\label{26b}
\begin{gathered}
  J_2 (x,t) = e^{ - i\frac{{\omega _s }}
{{v_g }}\left( {x + v_g t} \right)} \left( {2\pi  + i\,ci(\omega
_s \frac{{\left| x \right|}} {{v_g }}) + \,si(\omega _s
\frac{{\left| x \right|}}
{{v_g }})} \right. \hfill \\
  \left. { - i\,ci(\omega _s \frac{{x + v_g t}}
{{v_g }}) + si(\omega _s \frac{{x + v_g t}}
{{v_g }})} \right) \hfill \\
\end{gathered}
\end{equation}
where $x<0$, $x+v_gt>0$.

Therefore, the backscattered field can be written in the following
form:
\begin{widetext}
\begin{equation}\label{27}
\begin{gathered}
  u\left( {x,t} \right)/A = R(\omega_S)e^{ - i\frac{{\omega _s }}
{{v_g }}\left( {x + v_g t} \right)}  \hfill \\
   + \frac{iR(\omega_S)}{{2\pi }}e^{ -i\frac{{\Omega }} {{v_g }}(x + v_g t)}e^{ -\frac{{ \Gamma/2 }} {{v_g }}(x + v_g t)}
   \left( {   E_1\left( {  i\frac{{\left| x
\right|}} {{v_g }}\tilde \Omega } \right) + 2\pi i  -  E_1\left(
{-i\frac{{x + v_g t}}
{{v_g }}\tilde \Omega } \right)} \right) \hfill \\
   +\frac{R(\omega_S)}
{{2\pi }}e^{ - i\frac{{\omega _s }} {{v_g }}\left( {x + v_g t}
\right)} \left( {i\,ci(\omega _s \frac{{\left| x \right|}} {{v_g
}}) + \,si(\omega _s \frac{{\left| x \right|}} {{v_g }}) -
i\,ci(\omega _s \frac{{x + v_g t}} {{v_g }}) + si(\omega _s
\frac{{x + v_g t}}
{{v_g }})} \right) \hfill \\
\end{gathered}
\end{equation}

where $x<0$, $x+v_gt>0$.
\end{widetext}
Here, as in the case of the forward scattering there are three
terms in (\ref{27}), stationary solution, damping, and coherent
part of the scattered field.

\begin{figure}
  \includegraphics[width=8 cm]{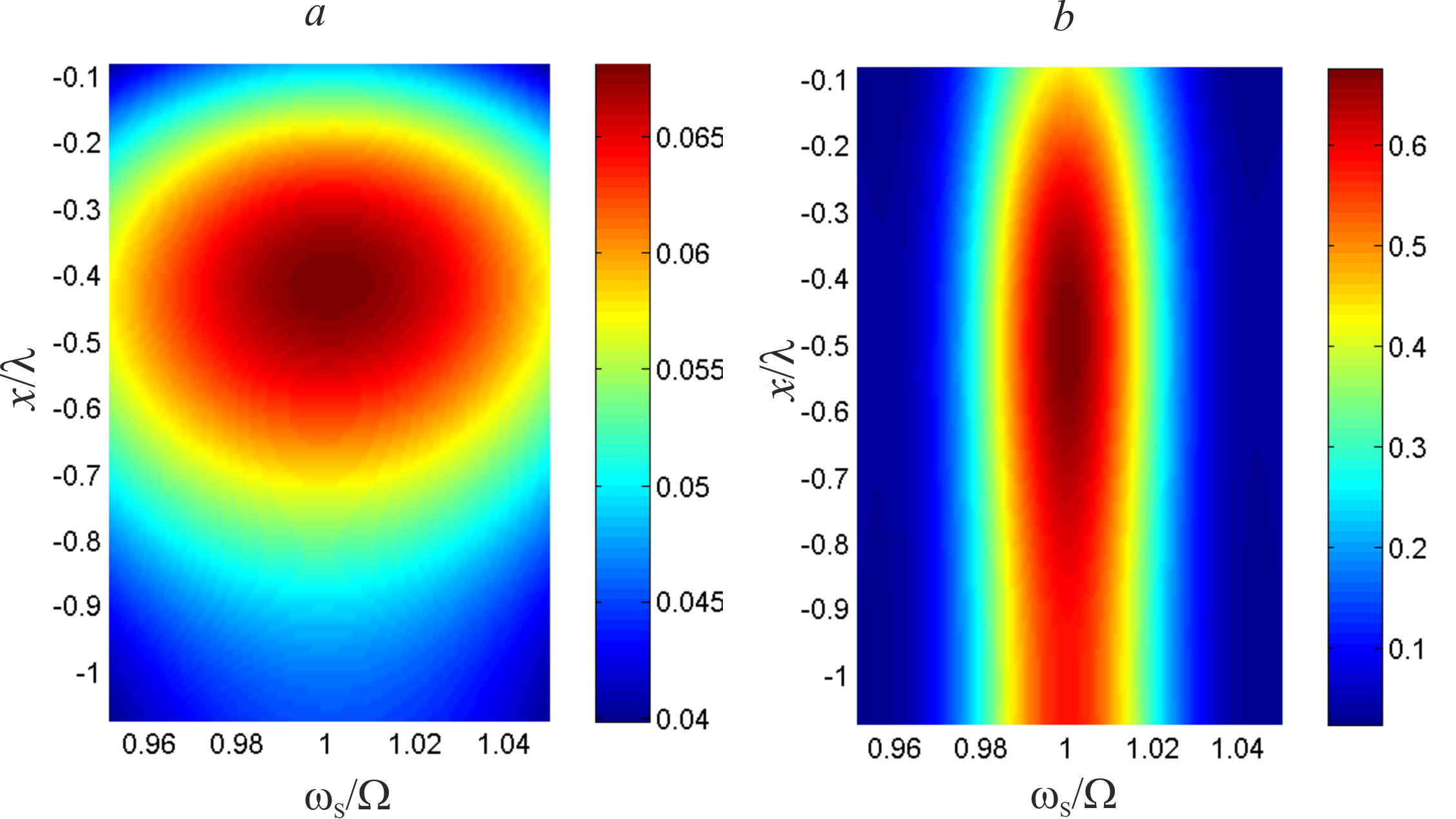}\\
  \caption{2D map of the reflectance calculated from (\ref{27}) for (a) $t=1 $ ns
 and (b) $t=5 $ ns. The color bar shows the value
  $|u(\omega_S,x,t)/A|^2$. $\Gamma/2\pi=0.01$ GHz, $\Omega/2\pi=5$ GHz, $\lambda=6$ cm.}\label{Fig2B}
\end{figure}

 Two dimensional maps of the reflectance
$|u(\omega_S,x,t)/A|^2$ calculated from (\ref{27}) for $t=1 $ ns,
$t=5 $ ns are shown in Fig. \ref{Fig2B}.

The first terms in (\ref{23a}) and (\ref{27}) are just the
transmission and reflection amplitudes from the stationary theory.
The second lines describe the field generated by spontaneous
emission of excited qubit. This field dies out as the time tends
to infinity. The third lines  are the transmitted and reflected
travelling waves which originate from the interaction of a qubit
with the incident photon.

\begin{figure}
  \includegraphics[width=8 cm]{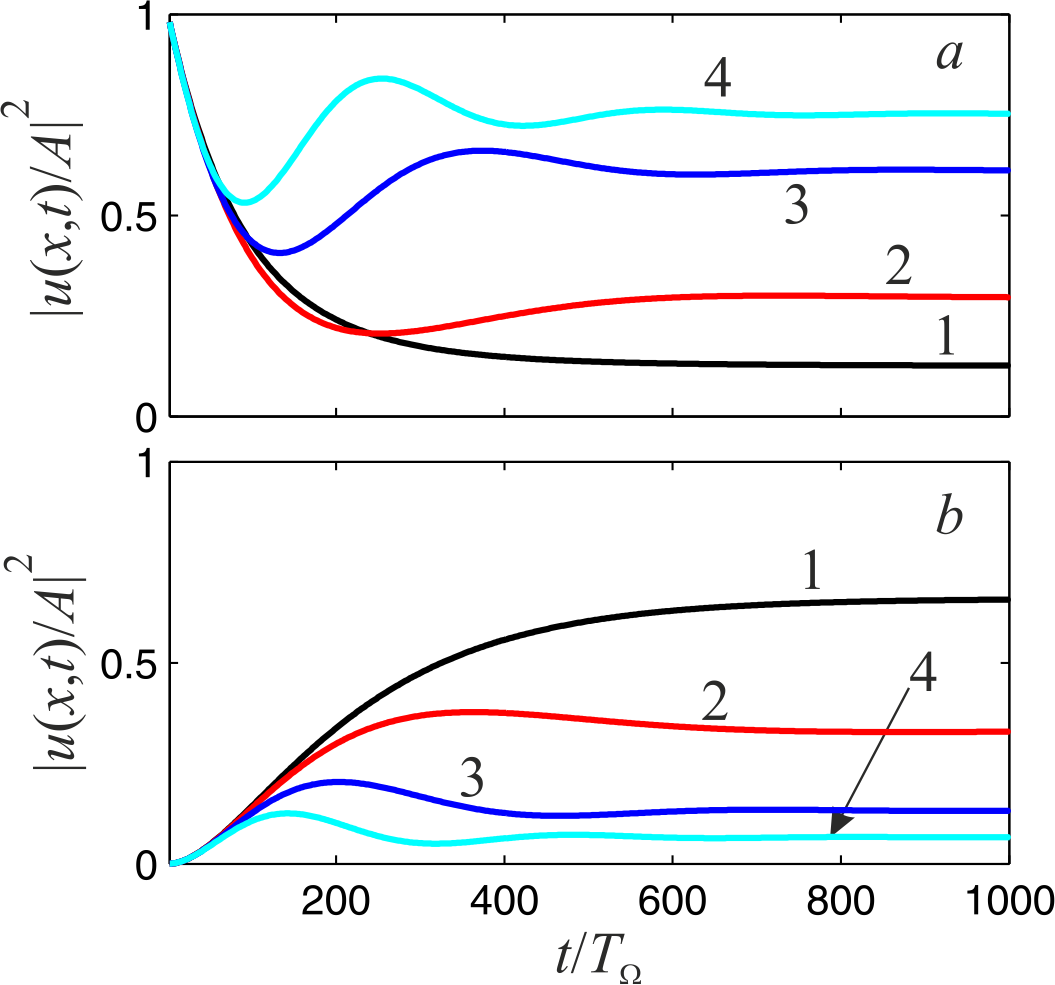}\\
  \caption{Dependence of $|u(x_0,t\geq t_0)/A|^2$ ($|x_0|=1$ mm, $t_0=10$ ps) for different
  frequencies: (1) $\omega_S=\Omega$; (2) $\omega_S=\Omega+0.5\Gamma$;
  (3) $\omega_S=\Omega+\Gamma$; (4) $\omega_S=\Omega+1.5\Gamma$; (a)
  transmittance (\ref{23a}), (b) reflectance (\ref{27}). $\Gamma/2\pi=0.01$ GHz,
  $\Omega/2\pi=5$ GHz, $T_\Omega=2\pi/\Omega$.}
   \label{Fig9A}
\end{figure}

It is worth noting that the scattered fields (\ref{23a}) and
(\ref{27}) display oscillatory behavior in time as shown in
Fig.\ref{Fig9A}. These oscillations at the frequency
$\omega_S-\Omega$ originate from the interference between the
first and the second (spontaneous decay) terms in expressions
(\ref{23a}) and (\ref{27}). To avoid the infinity of
$ci(x\omega_S/v_g)$ at $x=0$ we start the calculations in
Fig.\ref{Fig9A} at $|x_0|=1$ mm distance from the qubit and at the
time $t_0=10$ ps that insures the required condition $|x_0|-v_g
t_0<0$.

There exists a deep analogy between time oscillations of our
scattered fields and those of the decay probability in the
dynamics of unstable quantum system \cite{Peshkin2014}. In both
cases the time oscillations originate from the effective (after
the averaging out the photon degrees of freedom) non-Hermitian
Hamiltonian.

\subsection{The scattered field at large time}

There are three time scales in our problem: $1/\Delta$,
$1/\Gamma$, and $1/\Omega$ where $1/\Delta\gg 1/\Gamma\gg
1/\Omega$. As was shown in (\ref{G4}) a weak excitation probe sets
the upper bound on the time at which our theory is valid, $t\ll
1/\Delta$. Therefore, we may safely satisfy the conditions $\Gamma
t\gg 1$, $\Omega t\gg 1$ which are necessary to study the
asymptotic of the transmitted (\ref{23a}) and reflected (\ref{27})
fields for sufficiently large time.

If the time is sufficiently large and $x$ is fixed we may
disregard the time dependent corrections in (\ref{23a}) and
(\ref{27}). In this case, we obtain from (\ref{23a}) the field
behind the qubit:

\begin{equation}\label{23b}
\begin{gathered}
  u\left( {x,t} \right)/A = T(\omega_S)e^{i\frac{{\omega _s }}
{{v_g }}\left( {x - v_g t} \right)}  \hfill \\
   - i\frac{R(\omega_S)}
{{2\pi }}e^{i\frac{{\omega _s }} {{v_g }}\left( {x - v_g t}
\right)} \left( {i\,ci(\omega _s \frac{x} {{v_g }}) + si(\omega _s
\frac{x}
{{v_g }})} \right) \hfill \\
\end{gathered}
\end{equation}
where $T(\omega_S)$ and $R(\omega_S)$ are transmission and
reflection amplitudes (\ref{I1}) and (\ref{I2}), respectively;
$x>0$, $x-v_gt<0$.

From (\ref{23b}) we see that for $\omega_S=\Omega$ the field at
finite distance behind the qubit is non-zero. However, as $x$
tends to infinity ($x\gg\lambda$) the last term in (\ref{23b})
disappears and we are left with the stationary transmission
amplitude.

Similar calculations from (\ref{27}) provides the field ahead of
the qubit:

\begin{equation}\label{28}
\begin{gathered}
  u\left( {x,t} \right)/A = R(\omega_S)e^{ - i\frac{{\omega _s }}
{{v_g }}\left( {x + v_g t} \right)}  \hfill \\
   + \frac{R(\omega_S)}
{{2\pi }}e^{ - i\frac{{\omega _s }} {{v_g }}\left( {x + v_g t}
\right)} \left( {i\,ci(\omega _s \frac{{\left| x \right|}} {{v_g
}}) + \,si(\omega _s \frac{{\left| x \right|}}
{{v_g }})} \right) \hfill \\
\end{gathered}
\end{equation}
where $R(\omega_S)$ is the reflection amplitude (\ref{I2}); $x<0$,
$x+v_gt>0$. For sufficiently large time the field at finite
distance ahead of the qubit remains finite. However, as the $|x|$
tends to infinity ($|x|\gg\lambda$) the last term in (\ref{28})
disappears and we are left with the stationary reflection
amplitude.

\begin{figure}
  \includegraphics[width=8 cm]{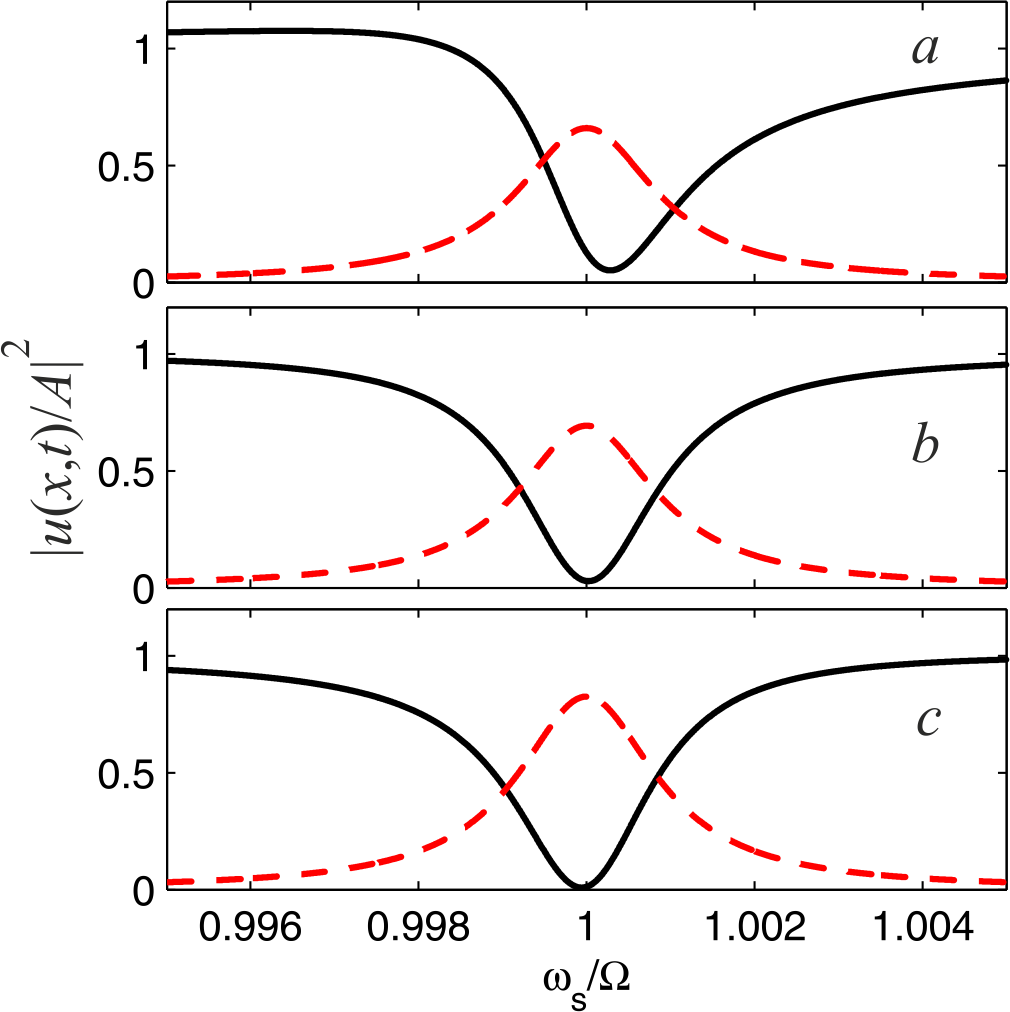}\\
  \caption{The dependence of the transmittance (\ref{23b}) (solid, black line) and
  reflectance (\ref{28}) (dashed, red line) on the photon
  frequency for different distances of the field point from the qubit. (a) $x=1$ mm,
  (b) $x=5$ mm, (c) $x=10$ mm, $\Gamma/2\pi=0.01$ GHz,
  $\Omega/2\pi=5$ GHz, $\lambda=6$ cm.}\label{Fig2}
\end{figure}

We investigate now how the scattered field (second terms in
(\ref{23b}) and (\ref{28})) influences the amplitude-frequency
curves (AFC) of transmitted and reflected signals. The dependence
of AFCs on the distance from qubit is shown in Fig.\ref{Fig2}.

\begin{figure}
  \includegraphics[width=8 cm]{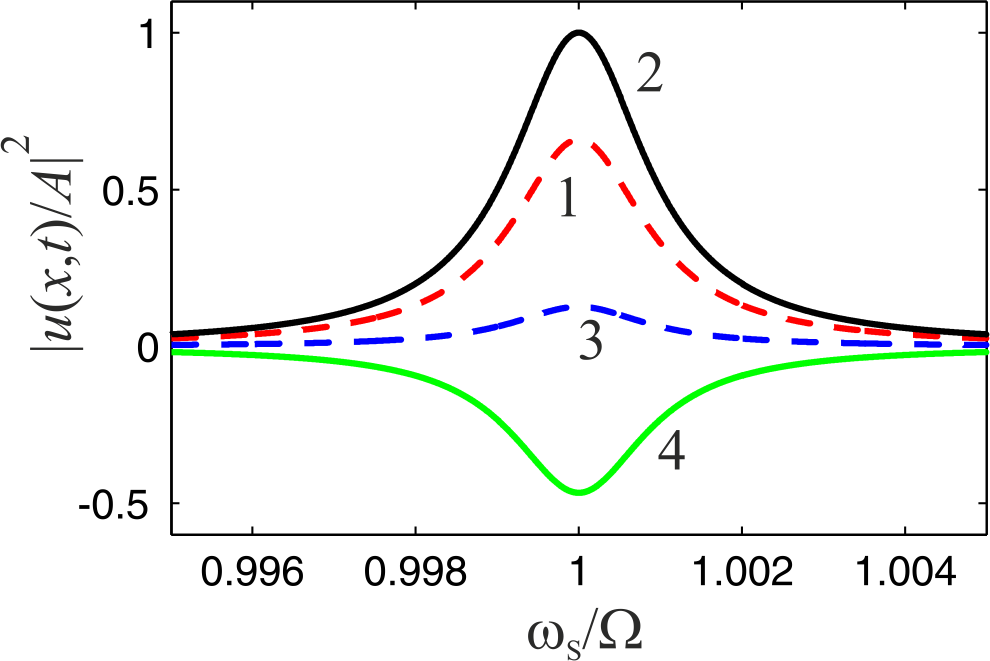}\\
  \caption{The influence of the interference on the reflectance (\ref{28})
  at 1 mm distance from the qubit.
  Dashed red line 1, the reflectance $|u(x,t)|^2/A^2$;
   solid black line 2, the reflectance $|R(\omega_S)|^2$ in the absence of
  interference; dashed blue line 3, the term $|z|^2$; solid green line 4, the
interference term $2|R(\omega_S)|^2Re(z)$. $\Gamma/2\pi=0.01$ GHz,
$\Omega/2\pi=5$ GHz.}\label{Fig6}
\end{figure}

We see that a clear asymmetry is observed at $x=1$ mm
($x\ll\lambda$) for transmitted AFC (Fig.\ref{Fig2}a). However,
for larger $x$ the asymmetry persists as well. We see in
Fig.\ref{Fig2}b and Fig.\ref{Fig2}c that the transmitted signal at
resonance ($\omega_S=\Omega$) is practically zero, while the
amplitude of the reflected signal at resonance is appreciably
smaller than unity. It can be attributed to the interference
between two terms in (\ref{28})). In fact, from (\ref{28}) we can
write the squared modulus of the reflected field as
$|u(x,t)|^2=A^2|R(\omega_S)^2|(1+z)|^2$ where $z$ is the term in
the round brackets in (\ref{28}). The influence of $z$  on the
reflected field at the 1 mm distance from the qubit is shown in
Fig.\ref{Fig6}.

As is seen from this figure, the contribution of the interference
term $2|R(\omega_S)|^2Re(z)$ is negative (the curve 4 in
Fig.\ref{Fig6}) and is significant. If the interference term in
(\ref{28}) is neglected we obtain $|R(\omega_S)|^2$ for
transmitted signal (curve 2 in Fig.\ref{Fig6}).

\begin{figure}
  \includegraphics[width=8 cm]{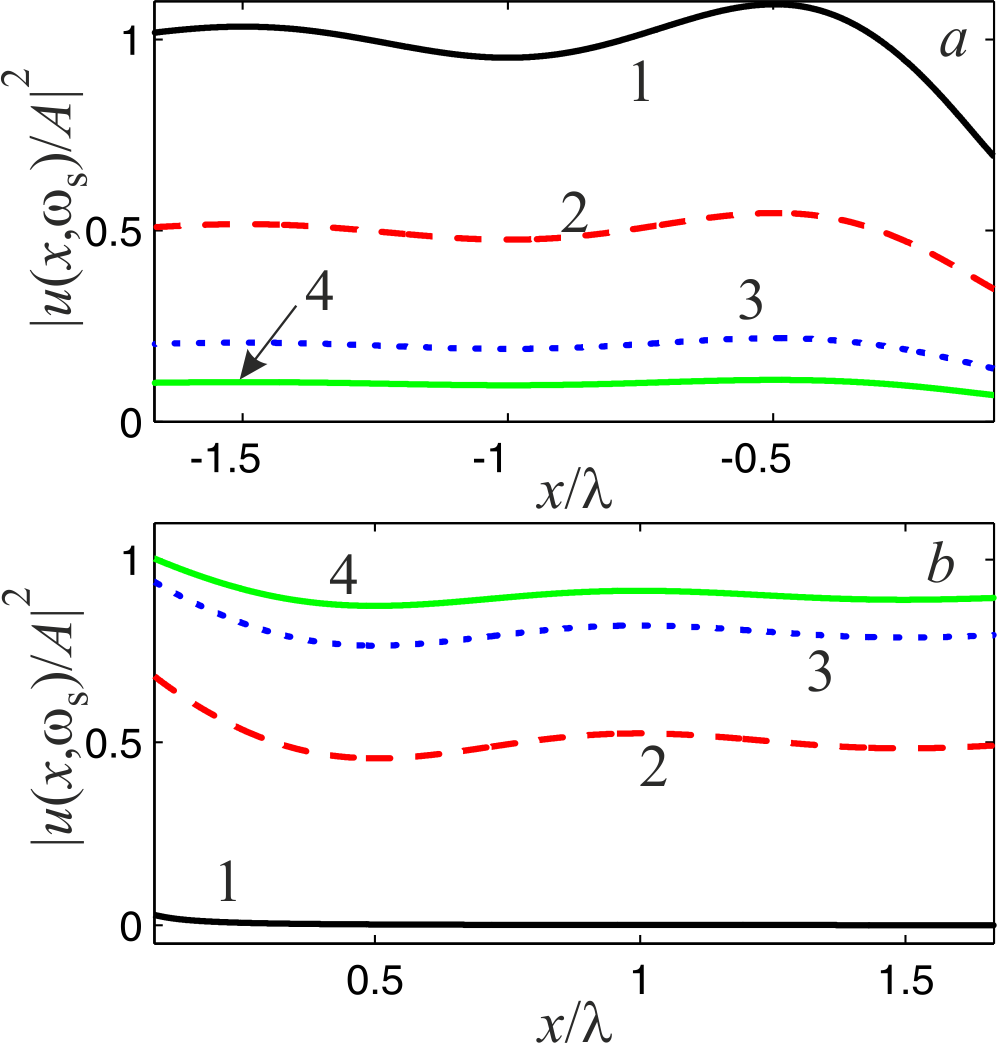}\\
  \caption{Spatial dependence of (a) reflectance (\ref{28}) and  (b)
  transmittance (\ref{23b})
  for off-resonant conditions. Solid black line
  (1), $\omega_S=\Omega$;
  dashed red line (2), $\omega_S=\Omega+0.5\Gamma$; dotted blue line (3),
 $\omega_S=\Omega+\Gamma$; solid green line (4), $\omega_S=\Omega+1.5\Gamma$.
 $\Gamma/2\pi=0.01$ GHz, $\Omega/2\pi=5$ GHz, $\lambda=6$ cm.}\label{Fig3}
\end{figure}

For off-resonant conditions, the interference effects persist both
for reflected and transmitted fields. The influence of these
effects on spatial dependence of the scattered fields is shown in
Fig.\ref{Fig3} for $\omega_S=\Omega, \Omega+0.5\Gamma,
\Omega+\Gamma, \Omega+1.5\Gamma$.

\begin{figure}
  \includegraphics[width=8 cm]{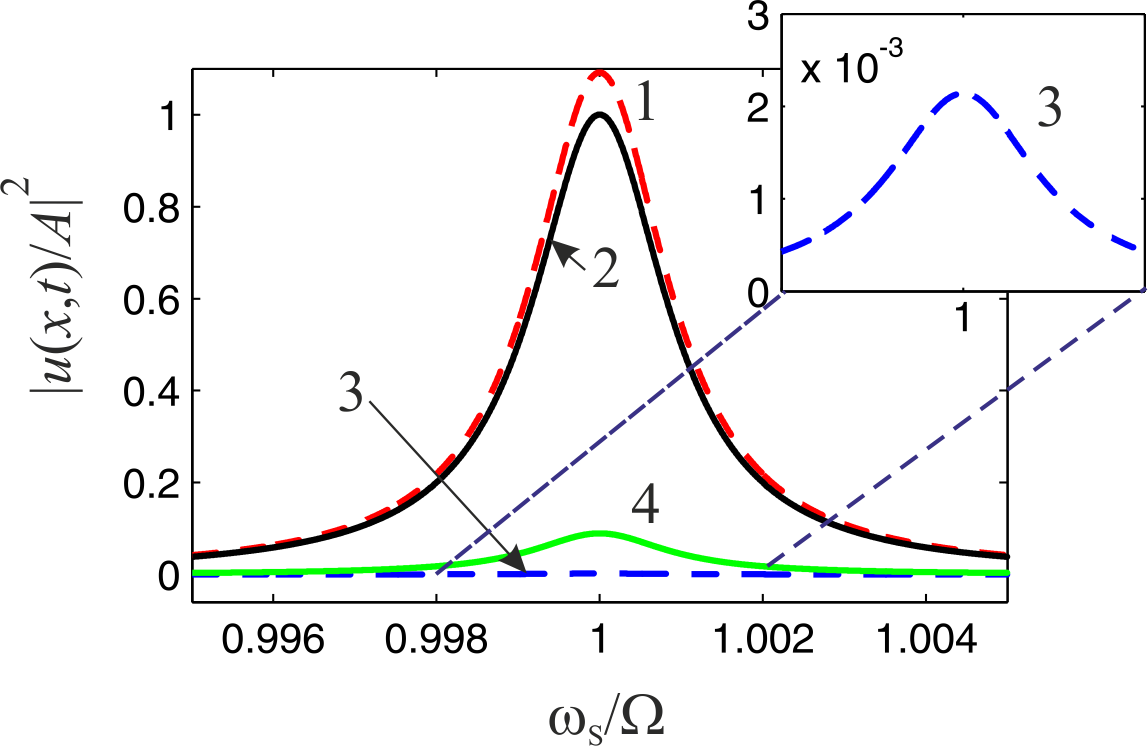}\\
  \caption{The influence of the interference on the reflectance (\ref{28})
  at $x=\lambda/2$ distance from the qubit. Dashed red line 1,
  the reflectance $|u(x,t)|^2/A^2$; solid black line 2, the reflectance $|R(\omega_S)|^2$
  in the absence of the interference; dashed blue line 3, the term $|z|^2$;
  solid green line 4, the interference term $2|R(\omega_S)|^2Re(z)$.
  $\Gamma/2\pi=0.01$ GHz, $\Omega/2\pi=5$ GHz, $\lambda=6$ cm.}\label{Fig6ampl}
\end{figure}

From Fig. \ref{Fig3}a we see that the reflectance at $x=\lambda/2$
is larger that 1 (see our comment below Fig.\ref{Fig2A}b). This
amplification can be explained by a constructive interference
between incident and reflected waves as is shown in
Fig.\ref{Fig6ampl}. In this case the interference term
$2|R(\omega_S)|^2Re(z)$ (solid green line 4 in Fig.\ref{Fig6ampl})
is positive (compare it with the line 4 in Fig.\ref{Fig6}) which
gives rise to a small amplification of the reflected field in the
resonance region.

 In principle, the interference effects can persist over
relatively long distance. As an example we calculate from
(\ref{23b}) and (\ref{28}) the transmittance and reflectance for
off-resonant frequency $\omega_1=\Omega+0.5\Gamma$.

\begin{equation}\label{trns}
\left| {\frac{{u(x,t)}} {A}} \right|^2  = \frac{1} {2}\left| {1 -
\frac{1} {{2\pi }}\left( {ici(\alpha ) + si(\alpha )} \right)}
\right|^2
\end{equation}
where $\alpha=\omega_1 x/v_g$, $x>0$, $x-v_gt<0$.

\begin{equation}\label{rfl}
\left| {\frac{{u(x,t)}} {A}} \right|^2  = \frac{1} {2}\left| {1 +
\frac{1} {{2\pi }}\left( {ici(\alpha ) + si(\alpha )} \right)}
\right|^2
\end{equation}
where $\alpha=\omega_1 |x|/v_g$, $x<0$, $x+v_gt>0$.

\begin{figure}[h]
  \includegraphics[width=8 cm]{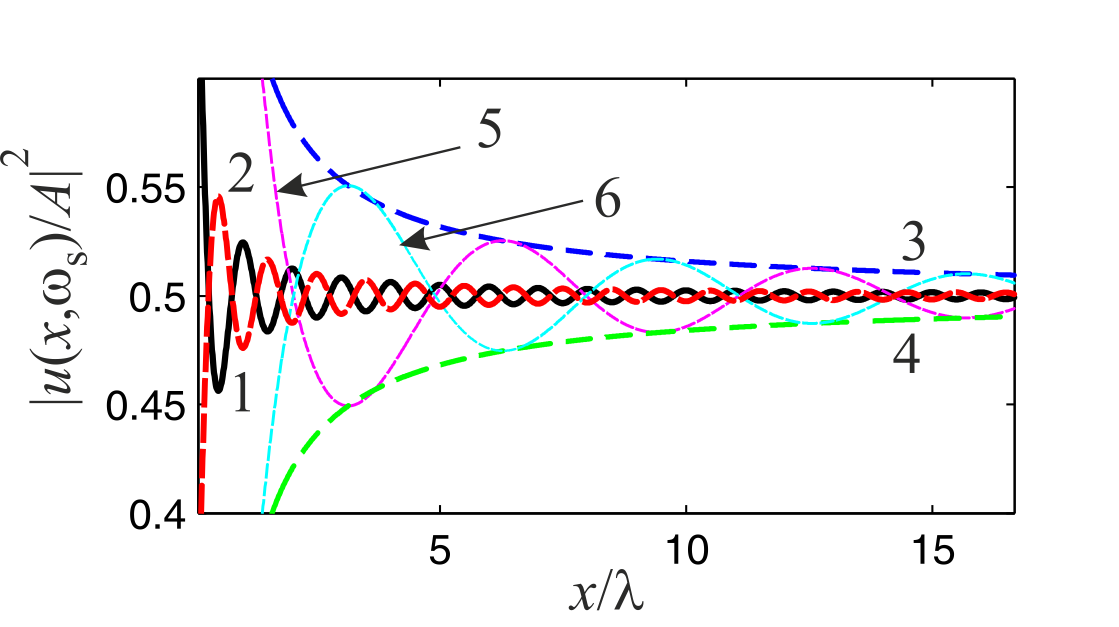}\\
  \caption{Spatial dependence of  transmittance, solid black  line 1 and
   reflectance,  dashed red line 2
  for off-resonant condition $\omega_S=\Omega+0.5\Gamma$,
  calculated from (\ref{trns}) and (\ref{rfl}), respectively.
  Solid magenta line 5 and solid cyan line 6 are the asymptotic
  behavior of transmittance and reflectance, respectively. Dashed
  blue 3 and green 4 lines are corresponding envelopes.
  $\Gamma/2\pi=0.01$ GHz, $\Omega/2\pi=5$ GHz, $\lambda=6$ cm.}\label{Fig4}
\end{figure}

The behavior of these quantities at the distance comparable to the
photon wavelength follows from the asymptotic of sine and cosine
integrals for large arguments (\ref{29}). The asymptotic behavior
of interference effects calculated from (\ref{trns}) and
(\ref{rfl}) is shown in Fig.\ref{Fig4}. The envelopes (lines 3 and
4 in Fig.\ref{Fig4}) scale as $\cos(2\pi x/\lambda)/(2\pi
x/\lambda)$.

\subsection{Asymptotic of the scattered field}

The behavior of scattered fields at large $x$ and $t$ follows from
the asymptotic of exponential integral function, sine integral,
and cosine integral  \cite{Grad2007, Jahnke}.

\begin{equation}\label{29}
    si(x)\approx -\frac{\cos(x)}{x}-\frac{\sin(x)}{x^2};
    \quad ci(x)\approx \frac{\sin(x)}{x}-\frac{\cos(x)}{x^2}
\end{equation}
where $x\gg 1$.
\begin{equation}\label{30}
    E_1(z)\approx \frac{e^{-z}}{z}\left(1-\frac{1}{z}\right)
\end{equation}
where $|z|\gg 1$.

With the aid of these approximations we obtain from (\ref{23a})
the asymptotic expression for forward scattering field.
\begin{widetext}
\begin{equation}\label{31}
\begin{gathered}
  u\left( {x,t} \right)/A = T(\omega_S)e^{i\frac{{\omega _s }}
{{v_g }}\left( {x - v_g t} \right)}  + \frac{{R(\omega_S)}} {{2\pi
}}\left( \frac{{v_g }} {{x\tilde \Omega }}{e^{ - i(\Omega  -
i\Gamma/2 )\;t}  - 2\pi e^{i\frac{{(\Omega  - i\Gamma/2 )}} {{v_g
}}(x - v_g t)}  + \frac{{v_g }}
{{\left| {x - v_g t} \right|\tilde \Omega }}} \right) \hfill \\
   - \frac{{R(\omega_S)}}
{{2\pi }}e^{i\frac{{\omega _s }} {{v_g }}\left( {x - v_g t}
\right)} \left( {\frac{{v_g }} {{\omega _s x}}e^{ - i\frac{{\omega
_S x}} {{v_g }}}  + \,\frac{{v_g }} {{\omega _s \left| {x - v_g t}
\right|}}e^{i\frac{{\omega _S \left| {x - v_g t} \right|}}
{{v_g }}} } \right) \hfill \\
\end{gathered}
\end{equation}
where $x>0$, $x-v_gt<0$, $v_g/\omega_Sx\ll 1$,
$v_g/|x-v_gt|\omega_S\ll 1$, $v_g/\widetilde{\Omega} x\ll 1$,
$v_g/|x-v_gt|\widetilde{\Omega}\ll 1$.

The asymptotic expression for backward scattering field reads:
\begin{equation}\label{32}
\begin{gathered}
  u\left( {x,t} \right)/A = R(\omega_S)e^{ - i\frac{{\omega _s }}
{{v_g }}\left( {x + v_g t} \right)}  + \frac{{R(\omega_S)}} {{2\pi
}}\left(  \frac{{v_g }} {{\left| x \right|\tilde \Omega }}{e^{ -
i(\Omega  - i\Gamma/2 )\;t} - 2\pi e^{ - i(\Omega  - i\Gamma/2
)\frac{{(x + v_g t)}} {{v_g }}}  + \frac{{v_g }}
{{(x + v_g t)\tilde \Omega }}} \right) \hfill \\
   - \frac{{R(\omega_S)}}
{{2\pi }}e^{ - i\frac{{\omega _s }} {{v_g }}\left( {x + v_g t}
\right)} \left( {e^{ - i\omega _s \frac{{\left| x \right|}} {{v_g
}}} \frac{{v_g }} {{\omega _S \left| x \right|}} + e^{i\omega _s
\frac{{x + v_g t}} {{v_g }}} \frac{{v_g }}
{{\omega _S (x + v_g t)}}} \right) \hfill \\
\end{gathered}
\end{equation}

where $x<0$, $x+v_gt>0$, $v_g/\omega_S|x|\ll 1$,
$v_g/(x+v_gt)\omega_S\ll 1$, $v_g/\widetilde{\Omega} |x|\ll 1$,
$v_g/|x+v_gt|\widetilde{\Omega}\ll 1$.
\end{widetext}

We see from (\ref{31}) and (\ref{32}) that the approach to the
stationary limit is very slow. The scattered field decreases as
$x^{-1}$ and $t^{-1}$ as the distance from the qubit and the time
after the interaction increase.

\section{Summary}
In summary, we have developed the time-dependent theory of the
scattering of a narrow single-photon Gaussian pulse on a qubit
embedded in 1D open waveguide. For a weak power of incident pulse
we have obtained explicit analytical expressions for the
transmitted and reflected fields, their spatial and time
dependence. We show that the scattered field consists of two
parts: a damping part which represents a spontaneous decay of the
excited qubit and a coherent, lossless part. The plain wave
solution for transmission and reflection amplitudes which are well
known from the stationary photon transport follow from our theory
as the limiting case when both the distance from the qubit and the
time after the scattering tend to infinity.

Even though our treatment can be applied to a real two-level atom,
we consider in our paper an artificial two-level atom, a
superconducting qubit operating at microwave frequencies at GHz
range. For our calculations we take qubit frequency
$\Omega/2\pi=5$ GHz which corresponds to wavelength $\lambda=6$
cm. Our calculations show that spatial effects can persist on the
scale of several $\lambda$'s (see Fig. \ref{Fig4}).  For on-chip
realization this length is not small compared with the dimensions
of a superconducting qubit (typically several microns). The power
of microwave signal is so low that the use of linear amplifiers
for the detection of the qubit signal is a common practice. The
current opportunity for on-chip realization of superconducting
qubit with associated circuitry allows for the placement of the
amplifier within the order of the wavelength from the qubit.
Therefore, in microwave range the near-field effects can in
principle be detectable.

We believe that the results obtained in this paper may have some
practical applications in quantum information technologies
including single-photon detection in a microwave domain as well as
the optimization of the readout of a qubit's quantum state.

\begin{acknowledgments}
Ya. S. G. thanks V. Kurin who attracted the author's attention to
the problem considered in the present paper.

The work is supported by the Ministry of Science and Higher
Education of Russian Federation under the project FSUN-2020-0004.
\end{acknowledgments}

\appendix
\section{Derivation of equation (\ref{10})}

The substitution of equations (\ref{6}) and (\ref{7}) in (\ref{5})
yields

\begin{equation}\label{A1}
\begin{gathered}
  \frac{{d\beta }}
{{dt}} =  - i\sum\limits_{k} {} g_k^{} \gamma _k(0)e^{ - i(\omega _k  - \Omega )t}  \hfill \\
 - \sum\limits_{k} {} g_k^2 \int\limits_0^t {} \beta (\tau)e^{ - i(\omega _k  - \Omega )(t - \tau)} d\tau \hfill \\
\end{gathered}
\end{equation}

In accordance with Wigner-Weisskopf approximation the quantity
$\beta{(t)}$ under
 the integrals in (\ref{A1}) is assumed to be  a slow function of time
as compared to that of the exponents. Therefore, for times
$\tau\ll t$ the integrand oscillates very rapidly and there is no
significant contribution to the value of the integral. The most
dominant contribution originates from times $\tau \approx t$. We
therefore evaluate $\beta{(t)}$ at the actual time $t$ and move it
out of the integrand. In this limit, the decay becomes a
memoryless process (Markov process).

\begin{equation}\label{A2}
\frac{{d\beta }} {{dt}} =  - i\sum\limits_k {} g_k^{} \gamma _k
(0)e^{ - i(\omega _k  - \Omega )t}  - \beta (t)\sum\limits_k {}
g_k^2 I_k (\Omega ,t)
\end{equation}
where
\begin{equation}\label{A3}
I_k (\Omega ,t) = \int\limits_0^t {e^{ - i(\omega _k  - \Omega )(t
- \tau )} d\tau }  = \int\limits_0^t {e^{ - i(\omega _k  - \Omega
)\tau } d\tau }
\end{equation}

 To evaluate this integral we extend the upper integration
limit to infinity since there is no significant contribution for $
\tau>> t$.  Therefore, we obtain:
\begin{equation}\label{A4}
I_k (\Omega ,t) \approx \int\limits_0^\infty  {e^{ - i(\omega _k -
\Omega )\tau } d\tau }  = \pi \delta (\omega _k  - \Omega ) -
iP.v.\left( {\frac{1} {{\omega _k  - \Omega }}} \right)
\end{equation}

where $P.v.$ represents the Cauchy principal value, which leads to
a frequency shift. In what follows, we do not write explicitly
this shift, which is assumed to be included in the qubit
frequency.

Therefore, the equation (\ref{A2}) can be rewritten as follows:
\begin{equation}\label{A5}
\frac{{d\beta }} {{dt}} =  - i\sum\limits_k {} g_k^{} \gamma _k
(0)e^{ - i(\omega _k  - \Omega )t}  - \frac{\Gamma }{2}\beta (t)
\end{equation}

where $\Gamma$ is the rate of spontaneous emission into waveguide
modes, which is given by the Fermi golden rule:
\begin{equation}\label{A6}
\Gamma  = 2\pi \sum\limits_k {g_k^2 \delta (\omega _k  - \Omega )}
\end{equation}

For 1D case the summation over $k$ is replaced by the integration.

\begin{equation}\label{A7}
\sum\limits_k {}  \Rightarrow \frac{L}{{2\pi }}\int\limits_{ -
\infty }^\infty  {dk}  = \frac{L}{{\pi }}\int\limits_0^\infty
{d\left| k \right|}  = \frac{{L}}{{\pi \upsilon _g
}}\int\limits_0^\infty  {d\omega _k }
\end{equation}

where we take a linear frequency dispersion $\omega_k=v_g|k|$ well
above the cutoff frequency of a waveguide.

Applying this prescription to (\ref{A6}) we express the
 coupling $g_\Omega$ at the qubit resonance frequency
$\Omega$ ($g_\Omega\equiv g_0$) in terms of the rate of
spontaneous emission $\Gamma$:

\begin{equation}\label{A8}
g_0  = \left( {\frac{{v_g \Gamma }} {{2L}}} \right)^{1/2}
\end{equation}

The first term in right hand side of equation (\ref{A5}) then
takes the form:

\begin{equation}\label{A9}
\begin{gathered}
   - i\sum\limits_k {} g_k^{} \gamma _k (0)e^{ - i(\omega _k  - \Omega )t}  \hfill \\
   =  - ig_0\frac{L}
{{\pi }}\int\limits_{0}^{  \infty }  \gamma _k
(0)e^{ - i(\omega _k  - \Omega )t} d\left| k \right|  \hfill \\
   =  - ig_0\frac{L}
{{\sqrt{v_g} \pi }}\int\limits_0^{  \infty }  \gamma _0 (\omega  )e^{ - i(\omega  - \Omega )t} d\omega  \hfill \\
   =  - i\sqrt {\frac{{\Gamma L}}
{{2\pi ^2  }}} \int\limits_0^{  \infty } {\gamma _0 (\omega )e^{ - i(\omega  - \Omega )t} d\omega }  \hfill \\
\end{gathered}
\end{equation}

where $\gamma_k(0)$ and $\gamma_0(\omega)$ are the initial
Gaussian packets in $k$ space (\ref{G1}) and frequency domain
(\ref{G2}), respectively.

The coupling constant $g_k\equiv g_\omega$ in the first line of
(\ref{A9}) is a slowly varying function of $\omega$ around the
qubit frequency $\Omega$, therefore, it can be taken out of the
integral in the second line of (\ref{A9}).

Therefore, for equation (\ref{A5}) we finally obtain the equation
(\ref{10}) from the main text:
\begin{equation}\label{A10}
\frac{{d\beta }} {{dt}} =  - i\sqrt {\frac{{\Gamma L}} {{2\pi ^2
}}} \int\limits_0^{  \infty } {\gamma _0 (\omega )e^{ - i(\omega -
\Omega )t} d\omega }  - \frac{\Gamma}{2} \beta (t)
\end{equation}

\section{Derivation of equations (\ref{21}), (\ref{22})}
\subsection{Calculation of $I_1(x,t)$}

The quantity $I_1(x,t)$ in (\ref{20a}) consists of two terms:
$I_1(x,t)=A(x,t)+B(x,t)$ where

\begin{equation}\label{B1}
\begin{gathered}
  A(x,t) = \int\limits_0^\infty  {} \frac{{e^{i\left( {\omega  - \widetilde{\Omega}  } \right)\;t} e^{i\frac{\omega }
{{v_g }}\left( {x - v_g t} \right)} }}
{{\left( {\omega  - \widetilde{\Omega} } \right)}}d\omega  \hfill \\
   = e^{-i { \widetilde{\Omega} }\;t} \int\limits_0^\infty  {} \frac{{e^{i\frac{\omega }
{{v_g }}x} }}
{{\left( {\omega  - \widetilde{\Omega} } \right)}}d\omega  \hfill \\
\end{gathered}
\end{equation}

\begin{equation}\label{B2}
B(x,t) =  - \int\limits_0^\infty  {} \frac{{e^{i\frac{\omega }
{{v_g }}\left( {x - v_g t} \right)} }} {{\left( {\omega  -
\widetilde{\Omega}} \right)}}d\omega
\end{equation}

\begin{figure}
  \includegraphics[width=8 cm, height=6 cm, keepaspectratio]{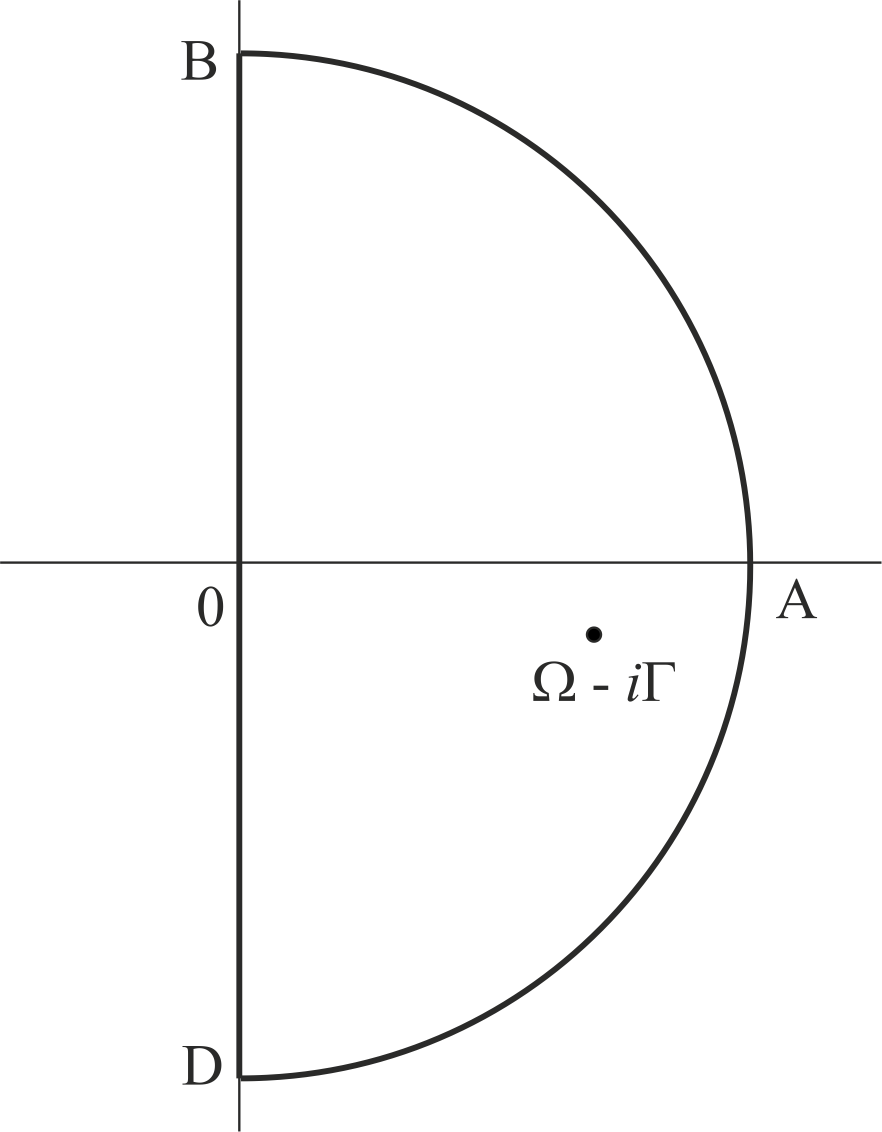}\\
  \caption{Plane of the complex $\omega$.}\label{Fig1}
\end{figure}

In the  plane of complex $\omega$ the only pole lies in the lower
part of the plane as shown in Fig.\ref{Fig1}. To calculate the
last integral in (\ref{B1}) for $x>0$ we take a closed contour
$C_1=OAB$ as shown in Fig.\ref{Fig1}. As there are no poles inside
this contour we obtain:
\begin{equation}\label{B3}
\begin{gathered}
  \oint\limits_{C_1 } {} \frac{{e^{i\frac{\omega }
{{v_g }}x} }} {{\left( {\omega  - \widetilde{\Omega} }
\right)}}d\omega  = 0 = \int\limits_0^\infty  {}
\frac{{e^{i\frac{\omega } {{v_g }}x} }}
{{\left( {\omega  - \widetilde{\Omega}} \right)}}d\omega  \hfill \\
   + \int\limits_{i\infty }^0 {} \frac{{e^{i\frac{\omega }
{{v_g }}x} }}
{{\left( {\omega  - \widetilde{\Omega}} \right)}}d\omega  \hfill \\
\end{gathered}
\end{equation}

\begin{equation}\label{B4}
\int\limits_0^\infty  {} \frac{{e^{i\frac{\omega } {{v_g }}x} }}
{{\left( {\omega  - \tilde \Omega } \right)}}d\omega  =  -
\int\limits_{i\infty }^0 {} \frac{{e^{i\frac{\omega } {{v_g }}x}
}} {{\left( {\omega  - \tilde \Omega } \right)}}d\omega  =
\int\limits_0^{i\infty } {} \frac{{e^{i\frac{\omega } {{v_g }}x}
}} {{\left( {\omega  - \tilde \Omega } \right)}}d\omega
\end{equation}

The last integral in (\ref{B4}) can be expressed in terms of the
exponential integral function $E_1(z)$ (p.228, 5.1.4 in
\cite{Abram1964}):
\begin{equation}\label{B5}
\int\limits_0^{i\infty } {\frac{{e^{i\frac{\omega } {{v_g }}x} }}
{{\left( {\omega  - \tilde \Omega } \right)}}d\omega }  =
\int\limits_0^\infty  {\frac{{e^{ - \alpha t} }} {{\left( {t +
\beta } \right)}}dt}  =   e^{\alpha \beta } E_1(  \alpha \beta )
\end{equation}

where $\alpha=x/v_g, \beta=i\Omega+\Gamma$. Therefore, for
$A(x,t)$ we obtain

\begin{equation}\label{B6}
A(x,t) =  e^{ - i\tilde \Omega \;t} e^{i\frac{x} {{v_g }}\tilde
\Omega } E_1\left( { i\frac{x} {{v_g }}\tilde \Omega } \right);\;x
> 0
\end{equation}

For the calculation of $B(x,t)$ for $x-v_gt<0$
 we must take the contour $C_2=OAD$ in the lower part of the complex $\omega$
 plane as shown in Fig.\ref{Fig1}:
\begin{equation}\label{B7}
\begin{gathered}
  \oint\limits_{C_2} \frac{{e^{i\frac{\omega }
{{v_g }}(x - v_g t)} }} {{\left( {\omega  - \widetilde{\Omega} }
\right)}}d\omega  =  - 2\pi ie^{i\frac{{\left( {\widetilde{\Omega}
} \right)}}
{{v_g }}(x - v_g t)}  \hfill \\
   = \int\limits_0^\infty  {} \frac{{e^{i\frac{\omega }
{{v_g }}(x - v_g t)} }} {{\left( {\omega  - \widetilde{\Omega} }
\right)}}d\omega  + \int\limits_{ - i\infty }^0 {}
\frac{{e^{i\frac{\omega } {{v_g }}(x - v_g t)} }}
{{\left( {\omega  - \widetilde{\Omega}} \right)}}d\omega  \hfill \\
\end{gathered}
\end{equation}
From (\ref{B7}) we obtain
\begin{equation}\label{B8}
\int\limits_0^\infty  {\frac{{e^{i\frac{\omega } {{v_g }}(x - v_g
t)} }} {{\left( {\omega  - \tilde \Omega } \right)}}d\omega }  = -
2\pi ie^{i\frac{{\tilde \Omega }} {{v_g }}(x - v_g t)}  -
\int\limits_{ - i\infty }^0 {\frac{{e^{i\frac{\omega } {{v_g }}(x
- v_g t)} }} {{\left( {\omega  - \tilde \Omega } \right)}}d\omega
}
\end{equation}
The last integral in (\ref{B8}) can be calculated similar to
(\ref{B5})
\begin{equation}\label{B9}
\begin{gathered}
  \int\limits_{ - i\infty }^0 {} \frac{{e^{i\frac{\omega }
{{v_g }}(x - v_g t)} }} {{\left( {\omega  - \tilde \Omega }
\right)}}d\omega  =  - \int\limits_0^\infty  {} \frac{{e^{ -
\frac{{\left| {x - v_g t} \right|}} {{v_g }}s} }}
{{s - i\tilde \Omega }}ds \hfill \\
   = -e^{ - i\frac{{\left| {x - v_g t} \right|}}
{{v_g }}\tilde \Omega } E_1\left(- {i\frac{{\left| {x - v_g t}
\right|}}
{{v_g }}\tilde \Omega } \right) \hfill \\
\end{gathered}
\end{equation}

Therefore, for $B(x,t)$ we obtain:

\begin{equation}\label{B10}
B(x,t) = 2\pi ie^{i\frac{{\tilde \Omega }} {{v_g }}(x - v_g t)} -
e^{ - i\frac{{\left| {x - v_g t} \right|}} {{v_g }}\tilde \Omega }
E_1\left(- {i\frac{{\left| {x - v_g t} \right|}} {{v_g }}\tilde
\Omega } \right)\;
\end{equation}

Combining (\ref{B6}) and (\ref{B10}) we finally obtain
\begin{equation}\label{B11}
\begin{gathered}
  I_1 (x,t) =  e^{ - i\tilde \Omega \;t} e^{i\frac{x}
{{v_g }}\tilde \Omega } E_1\left( {  i\frac{x}
{{v_g }}\tilde \Omega } \right) \hfill \\
   + 2\pi ie^{i\frac{{\tilde \Omega }}
{{v_g }}(x - v_g t)}  - e^{ - i\frac{{\left| {x - v_g t} \right|}}
{{v_g }}\tilde \Omega } E_1\left(- {i\frac{{\left| {x - v_g t}
\right|}}
{{v_g }}\tilde \Omega } \right)\; \hfill \\
\end{gathered}
\end{equation}

\subsection{Calculation of $I_2(x,t)$}
We rewrite (\ref{20b}) as follows:
\begin{equation}\label{B12}
I_2 (x,t) = e^{i\frac{{\omega _s }} {{v_g }}\left( {x - v_g t}
\right)} \frac{1} {i}\int\limits_0^\infty  {} \frac{{e^{i\left(
{\omega  - \omega _s } \right)t}  - 1}} {{^{\left( {\omega  -
\omega _s } \right)} }}e^{i\frac{{\left( {\omega  - \omega _s }
\right)}} {{v_g }}\left( {x - v_g t} \right)} d\omega
\end{equation}

In the integrand of (\ref{B12}) we introduce new variables
$\omega-\omega_s=z$, $(x-v_gt)/v_g=T$. We then obtain:

\begin{equation}\label{B13}
\begin{gathered}
  \int\limits_0^\infty  {} \frac{{e^{i\left( {\omega  - \omega _s } \right)t}  - 1}}
{{^{\left( {\omega  - \omega _s } \right)} }}e^{i\frac{{\left(
{\omega  - \omega _s } \right)}}
{{v_g }}\left( {x - v_g t} \right)} d\omega  \hfill \\
   = \int\limits_{ - \omega _s }^\infty  {} \frac{{e^{iz\tau } }}
{{^z }}dz - \int\limits_{ - \omega _s }^\infty  {} \frac{{e^{izT}
}}
{{^z }}dz \hfill \\
\end{gathered}
\end{equation}
where $\tau=x/v_g$.

For the first integral in (\ref{B13}) we obtain:
\begin{equation}\label{B14}
\begin{gathered}
  \int\limits_{ - \omega _s }^\infty  {\frac{{e^{iz\tau } }}
{{^z }}d} z = \int\limits_{ - \omega _s }^\infty  {\frac{{\cos
z\tau }} {{^z }}dz}  + i\int\limits_{ - \omega _s }^\infty
{\frac{{\sin z\tau }}
{{^z }}dz}  \hfill \\
   = \int\limits_{ - \omega _S }^{\omega _S } {\frac{{\cos z\tau }}
{{^z }}d} z + \int\limits_{\omega _S }^\infty  {\frac{{\cos z\tau
}} {{^z }}dz}  + i\int\limits_{ - \omega _s }^\infty  {\frac{{\sin
z\tau }}
{{^z }}dz}  \hfill \\
   = \int\limits_{\omega _s }^\infty  {\frac{{\cos z\tau }}
{{^z }}dz}  + i\int\limits_{ - \omega _s }^\infty  {\frac{{\sin
z\tau }}
{{^z }}dz}  =  - ci(\omega _s \tau ) - i\,si( - \omega _s \tau ) \hfill \\
\end{gathered}
\end{equation}
where we introduced the sine and cosine integrals:
\begin{equation}\label{B15}
\begin{gathered}
  ci(\omega _s \tau ) =  - \int\limits_{\omega _s }^\infty  {} \frac{{\cos z\tau }}
{{^z }}dz \hfill \\
  si( - \omega _s \tau) =  - \int\limits_{ - \omega _s}^\infty  {} \frac{{\sin z\tau }}
{{^z }}dz \hfill \\
\end{gathered}
\end{equation}

Similar calculations for the second integral in (\ref{B13}) yield:
\begin{equation}\label{B16}
\int\limits_{ - \omega _s }^\infty  {} \frac{{e^{izT} }} {{^z }}dz
=  - ci(\omega _s T) - i\,si( - \omega _s T)
\end{equation}
where
\begin{equation}\label{B17}
\begin{gathered}
  ci(\omega _s T) =  - \int\limits_{\omega _s  }^\infty  {} \frac{{\cos z T}}
{{^z }}dz \hfill \\
  si( - \omega _s T) =  - \int\limits_{ - \omega _s  }^\infty  {} \frac{{\sin z T}}
{{^z }}dz \hfill \\
\end{gathered}
\end{equation}

Finally we obtain
\begin{equation}\label{B18}
\begin{gathered}
  I_2 (x,t) = e^{i\frac{{\omega _s }}
{{v_g }}\left( {x - v_g t} \right)}  \hfill \\
   \times \left( {i\,ci(\omega _s \tau ) - \,si( - \omega _s \tau ) - i\,ci(\omega _s T) + si( - \omega _s T)} \right) \hfill \\
\end{gathered}
\end{equation}
Next, we use the known property of sine integral \cite{Grad2007}:
\begin{equation}\label{B19}
si(y) + si( - y) =  - \pi
\end{equation}
and two relations which follow from (\ref{B17}) for $T<0$:
\begin{equation}\label{B20}
\begin{gathered}
  si(-\omega_S T) =  - si(-\omega_S |T|) \hfill \\
  ci(\omega_S T) = ci(\omega_S |T|) \hfill \\
\end{gathered}
\end{equation}

Therefore, for $I_2(x,t)$ (\ref{B18}) where $\tau>0$ and $T<0$, we
finally obtain:
\begin{equation}\label{B21}
\begin{gathered}
  I_2 (x,t) = e^{i\frac{{\omega _s }}
{{v_g }}\left( {x - v_g t} \right)} \left( {2\pi  + i\,ci(\omega _s \tau ) + si(\omega _s \tau )} \right. \hfill \\
  \left. { - i\,ci(\omega _s \left| T \right|) + si(\omega _s \left| T \right|)} \right) \hfill \\
\end{gathered}
\end{equation}
which is the equation (\ref{22}) from the main text.

\section{The influence of the probing power, decoherence rate, and the non radiative
 losses on the transmitted and reflected fields}

With account for probing power and all losses the reflection
coefficient can be expressed as \cite{Asta2010, Hoi2013}

\begin{equation}\label{C1}
R(\omega _S ) =  - \frac{\Gamma } {{2\gamma }}\frac{{1 + i\delta
\omega _S /\gamma }} {{1 + \left( {\delta \omega _S /\gamma }
\right)^2  + \Omega _R^2 /(\Gamma  + \Gamma _l )\gamma }}
\end{equation}

where $\delta\omega_S=\omega_S-\Omega$, $\Omega_R$ is the Rabi
oscillation frequency, the square of which is proportional to the
power, $P$ of incident wave,
$\gamma=\frac{\Gamma}{2}+\Gamma_\varphi+\frac{\Gamma_l}{2}$ is the
total decoherence rate where $\Gamma_\varphi$ is pure dephasing
and $\Gamma_l$ is the non-radiative intrinsic losses. The
transmission coefficient can be found from the relation $T=1+R$,
which holds for a single emitter \cite{Asta2010, Hoi2013}.
\begin{equation}\label{C2}
T(\omega _S ) = \frac{{1 + \left( {\delta \omega _S /\gamma }
\right)^2  - \frac{\Gamma } {{2\gamma }}\left( {1 + i\delta \omega
_S )/\gamma } \right) + \frac{\Omega _R^2 }{(\Gamma  + \Gamma _l
)\gamma }}} {{1 + \left( {\delta \omega _S /\gamma } \right)^2  +
\frac{\Omega _R^2}{(\Gamma  + \Gamma _l )\gamma}  }}
\end{equation}

For a probe power in the single-photon regime,
$\Omega_R\ll\Gamma$, we obtain from (\ref{C1}) and (\ref{C2}):
\begin{equation}\label{C3}
R(\omega _S ) = \frac{{ - i\frac{\Gamma } {2}}} {{\omega  - \Omega
+ i\left( {\frac{\Gamma } {2} + \Gamma _\varphi   + \frac{\Gamma
_l}{2} } \right)}}
\end{equation}

\begin{equation}\label{C4}
T(\omega _S ) = \frac{{\omega  - \Omega  + i\left( {\Gamma
_\varphi + \frac{{\Gamma _l }} {2}} \right)}} {{\omega  - \Omega
+ i\left( {\frac{\Gamma } {2} + \Gamma _\varphi   + \frac{{\Gamma
_l }} {2}} \right)}}
\end{equation}

The equations (\ref{C3}) and (\ref{C4}) coincide with (\ref{I1})
and (\ref{I2}) if we neglect pure dephasing and non-radiative
losses. However, as it follows from (\ref{C3}) and (\ref{C4}) the
pure dephasing and non-radiative losses can be included in the
framework of our treatment simply by the redifinition of  the
qubit's frequency $\Omega$,
$\Omega\rightarrow\Omega-i(\Gamma_\varphi+\Gamma_l/2)$.

The coupling of the qubit to a waveguide can be described by a
relevant quantity $\beta=\Gamma/2\gamma$. If we disregard
$\Gamma_\varphi$ and $\Gamma_l$ we obtain the critical coupling
$\beta=1$ which means at resonant frequency, $\omega_S=\Omega$ a
full extinction of transmitted signal, $|T(\Omega)|^2=0$ and a
complete reflection $|R(\Omega)|^2=1$. However, if we account for
dephasing and non radiative losses the full extinction of
transmitted field and complete reflection never happen.

\end{document}